# An Eclipsing Substellar Binary in a Young Triple System discovered by SPECULOOS

Amaury H.M.J. Triaud* ORCID 0000-0002-5510-8751
School of Physics & Astronomy, University of Birmingham, Edgbaston, Birmingham B15 2TT, United Kingdom

Adam J. Burgasser* ORCID 0000-0002-6523-9536
Center for Astrophysics and Space Sciences, University of California, San Diego, 9500 Gilman Drive, Mail Code 0424, La Jolla, CA 92093, USA

Artem Burdanov
Astrobiology Research Unit, Université de Liège, Allée du 6 août 19, Sart Tilman, 4000, Liège 1, Belgium

Vedad Kunovac Hodžić ORCID 0000-0001-9419-3736
School of Physics & Astronomy, University of Birmingham, Edgbaston, Birmingham B15 2TT, United Kingdom

Roi Alonso
Instituto de Astrofísica de Canarias, C/ Vía Láctea s/n, 38205 La Laguna, Tenerife, Spain

Daniella Bardalez Gagliuffi ORCID 0000-0001-8170-7072
Department of Astrophysics, American Museum of Natural History, 200 Central Park West at 81st St., New York, NY 10024, USA

Laetitia Delrez
Cavendish Laboratory, J J Thomson Avenue, Cambridge, CB3 0HE, UK

Brice-Olivier Demory ORCID 0000-0002-9355-5165
University of Bern, Center for Space and Habitability, Gesellschaftsstrasse 6, 3012 Bern, Switzerland

Julien de Wit
Department of Earth, Atmospheric and Planetary Sciences, Massachusetts Institute of Technology, 77 Massachusetts Avenue, Cambridge, Massachusetts 02139, USA

Elsa Ducrot ORCID 0000-0002-7008-6888
Astrobiology Research Unit, Université de Liège, Allée du 6 août 19, Sart Tilman, 4000, Liège 1, Belgium

Frederic V. Hessman
Institut für Astrophysik, University of Göttingen, Friedrich-Hund-Platz 1, D-37077 Göttingen, Germany

Tim-Oliver Husser
Institut für Astrophysik, University of Göttingen, Friedrich-Hund-Platz 1, D-37077 Göttingen, Germany

Emmanuël Jehin ORCID 0000-0001-8923-488X
Space Sciences, Technologies and Astrophysics Research (STAR) Institute, Université de Liège, Belgium

Peter P. Pedersen
Cavendish Laboratory, J J Thomson Avenue, Cambridge, CB3 0HE, UK

Didier Queloz
Cavendish Laboratory, J J Thomson Avenue, Cambridge, CB3 0HE, UK

James McCormac ORCID 0000-0003-1631-4170
Department of Physics, University of Warwick, Gibbet Hill Road, Coventry CV4 7AL, UK

Catriona Murray
Cavendish Laboratory, J J Thomson Avenue, Cambridge, CB3 0HE, UK

Daniel Sebastian
Astrobiology Research Unit, Université de Liège, Allée du 6 août 19, Sart Tilman, 4000, Liège 1, Belgium

Samantha Thompson
Cavendish Laboratory, J J Thomson Avenue, Cambridge, CB3 0HE, UK

Valérie Van Grootel
Space Sciences, Technologies and Astrophysics Research (STAR) Institute, Université de Liège, Belgium

Michaël Gillon ORCID 0000-0003-1462-7739
Astrobiology Research Unit, Université de Liège, Allée du 6 août 19, Sart Tilman, 4000, Liège 1, Belgium

*corresponding authors

**Mass, radius, and age are three of the most fundamental parameters for celestial objects, enabling studies of the evolution and internal physics of stars, brown dwarfs, and planets. Brown dwarfs are hydrogen-rich objects that are unable to sustain core fusion reactions but are supported from collapse by electron degeneracy pressure [1]. As they age, brown dwarfs cool, reducing their radius and luminosity. Young exoplanets follow a similar behaviour. Brown dwarf evolutionary models are relied upon to infer the masses, radii and ages of these objects [2, 3]. Similar models are used to infer the mass and radius of directly imaged exoplanets [4]. Unfortunately, only sparse empirical mass, radius and age measurements are currently available, and the models remain mostly unvalidated. Double-line eclipsing binaries provide the most direct route for the absolute determination of the masses and radii of stars [5, 6, 7]. Here, we report the SPECULOOS discovery of 2M1510A, a nearby, eclipsing, double-line brown dwarf binary, with a widely-separated tertiary brown dwarf companion. We also find that the system is a member of the 45±5 Myr-old moving group, Argus [8, 9]. The system's age matches those of currently known directly-imaged exoplanets. 2M1510A provides an opportunity to benchmark evolutionary models of brown dwarfs and young planets. We find that widely-used evolutionary models [4] do reproduce the mass, radius and age of the binary components remarkably well, but overestimate the luminosity by up to 0.65 magnitudes, which could result in underestimated photometric masses for directly-imaged exoplanets and young field brown dwarfs by 20 to 35%.**

There is only one double-line eclipsing substellar binary currently known: 2MASS J05352184-0546085AB (hereafter 2M0535) [10, 11]. A member of the 1 Myr Orion Nebular Cluster, this system is a singular benchmark for evolutionary models of brown dwarfs, but its youth implies that component parameters could be affected by differential formation age, ongoing accretion, rapid rotation, and structural and surface effects induced by strong magnetic activity. Consequently, their parameters have proven difficult to reconcile with evolutionary models [12]. Masses and radii have also been determined for roughly a dozen brown dwarfs identified in transit around more massive stars [13, 14]. While their ages are relatively mature (> 1 Gyr), the significant scatter of these objects in the mass-radius plane suggests that their evolution may be influenced by their proximity to their host star [13]. Robust tests of substellar evolutionary models necessitate the identification of substellar double-line eclipsing binaries in clusters of various ages. 2M1510, the system we present here, is older than 2M0535, and has smaller radii that are intermediate between young and old brown dwarfs.

We have identified a double-line eclipsing binary system within a compact, fully substellar triple system, which is also a member of a nearby stellar association older than Orion. The brown dwarf binary 2MASSW J1510478-281817 (hereafter 2M1510A) and its visual companion 2MASS J15104761-2818234 (hereafter 2M1510B) were first identified in 2MASS data [16] as a resolved (separation 6.8 arcsec), unequal-brightness (Delta J = 1.17±0.04), M dwarf visual binary [15]. 2M1510A is known to exhibit H$\alpha$ emission [15] (see Methods). With an average *Gaia*-measured distance of 36.6±0.3 pc [17 – Bailer-Jones et al. 2018], the projected separation between 2M1510A and 2M1510B spans 250 au. 2M1510A has been identified as a kinematic member of the 45±5 Myr Argus moving group [8, 9], and we confirm membership for this component and 2M1510B with *Gaia* data and radial velocity measurements reported here (see Methods). In addition, newly-acquired low-resolution near-infrared spectra for both sources exhibit low surface gravity absorption features consistent with 10-100 Myr substellar objects (see Methods). Despite their unequal brightness, the near-infrared and red optical spectra of these sources have similar classifications, leading to the early proposition that 2M1510A might itself be "*an equal luminosity double*" [15]. We name the two components 2M1510Aa and 2M1510Ab, forming a tight inner binary, with 2M1510B as the tertiary.

Both sources were simultaneously monitored during commissioning observations of the SPECULOOS-South (Search for habitable Planets EClipsing ULtra-cOOl Stars) Observatory at Cerro Paranal, Chile, which consists of four robotic telescopes, each with a one-metre diameter primary mirror [18]. The main goal of the SPECULOOS-South Observatory is to identify rocky planets transiting the lowest-mass stars and brown dwarfs, such as those found in the TRAPPIST-1 system [19]. We began monitoring 2M1510A on 2017 July 18. On 2017 July 26, we detected a 4% drop in brightness over a 90-minute duration, with a profile consistent with a grazing eclipse (see Fig. 1a). We promptly obtained high-resolution, near-infrared spectra of both sources using the Near-InfraRed Spectrometer (NIRSPEC [20]) on the Keck II telescope on 2017 August 02. These data (see Methods) revealed 2M1510A to be a double-line spectroscopic binary with component radial velocity separation of 22.9±1.0 km/s indicating a relatively short orbital period. In the following months, we obtained twelve high-resolution red optical spectra of 2M1510A using the red arm of the UV-Visual Échelle Spectrometer (UVES [21]) on the Very Large Telescope UT2 (Kueyen), and another six Keck/NIRSPEC near-infrared spectra (Figure 1b and 1c). We extracted component radial velocities from these 19 spectra as detailed in Methods. The velocity components of 2M1510A, which are distinguishable by their slightly unequal relative fluxes, show a clear pattern of oscillation, consistent with two brown dwarfs in close orbit around a common centre of mass.

We jointly fit the primary and secondary radial velocities of 2M1510A with SPECULOOS photometry to a binary model, assuming Keplerian orbits. Our models are produced with *ellc* [22] and converged using the *emcee* sampler [23], in a tested python package named *amelie* [14]. We found a solution and estimated uncertainties using Markov Chain Monte Carlo techniques, as described in Methods. All orbital and physical parameters are provided in Table 1. The data are consistent with a pair of near-equal mass brown dwarfs: 2M1510Aa, with a mass of 0.0382±0.0027 $M_\odot$ (40.0 $M_{Jup}$), and 2M1510Ab, with a mass of 0.0375±0.0029 $M_\odot$ (39.3 $M_{Jup}$). The pair occupies a 20.902±0.006 day orbit, with a non-zero eccentricity e = 0.31±0.02 and an orbital inclination i = 88.5±0.1°. Our orbital solution determines that the observed grazing eclipse occurs when the primary passes in front of the secondary (secondary eclipse) near periastron. The primary eclipse, which occurs close to apoastron, is undetectable due to the tilt and semi-major axis of the orbit (a = 0.063±0.001 AU). This prevents measurement of individual component radii, but we can constrain the sum of the radii (see Methods) to be 0.315±0.016 $R_\odot$ (3.15 $R_{Jup}$). Since both masses are equivalent to within their uncertainties, we assume that each component has a radius equal to half of their sum. We note that the NIRSPEC and UVES spectra indicate a flux ratio $f_2/f_1$ = 0.83-0.88 between both components, which could originate either from a radius ratio of 0.91-0.94 (assuming equal temperatures, a temperature ratio of 0.95-0.97 (assuming equal radii), or a mixture of the two. Since we are unable to differentiate between these cases, we assume equal radii for our subsequent analysis.

We compare our measurements to evolutionary models [4] in Figures 2 and 3. We find that the masses and radii are independently compatible with the age of the Argus moving group, 45±5 Myr [9] to within 1σ. In Figure 3 we present mass tracks as a function of radius and age (3a), and luminosity and age (3b). Those same models also predict individual component magnitudes, which we find to be brighter than the observed absolute magnitudes (see Methods). Each component of 2M1510A appears fainter by 0.65±0.14, 0.48±0.13, and 0.40±0.13 in J, H, K magnitudes, respectively, than the model predictions. Our estimate of the bolometric luminosity is similarly below model predictions (Figure 3). This disagreement is consistent with prior measurements of young brown dwarfs and planetary mass objects. For example, [24] measure a luminosity of β Pic b that is 0.7 dex below hot-start model predictions for that source's mass and age, while young ultracool dwarfs are found to be both under- or over-luminous than models in color-magnitude diagrams by up to 1 magnitude (or equivalently too blue or too red; see [25]). Model magnitudes are often used to obtain mass estimates (or detection limits) for directly imaged planets [26]. Should the magnitude differences measured for 2M1510Aab and other young brown dwarfs extend to massive exoplanets, our results suggest that the latter masses can be systematically underestimated by 20-35%. Alternatively, if we assume that the dynamical masses and the bolometric luminosity are correct, the models imply that the age of the system is of order 70 Myr (see Methods). In this case the measured radii would become 1.09 larger than predicted by the models [4].

The formation of close binaries, particularly those with non-zero eccentricities remains an area of debate [27, 28]. One possible formation mechanism is the dynamic perturbation of a wide tertiary companion in a misaligned orbital plane. If the outer orbit is inclined by more than 40°, the inner pair can experience significant variations in eccentricity, approaching unity, through so-called Lidov-Kozai cycles [28, 29, 30]. When the inner pair is at periastron, tidal dissipation works to circularise the orbit and reduce its orbital semimajor axis. We investigated whether 2M1510A's parameters are compatible with Lidov-Kozai cycles (see Methods). We conclude that this is unlikely the case: 2M1510B's separation is too large, leading to Lidov-Kozai cycles that would greatly exceed the system's current age. The eccentricity we detect is therefore likely primordial, and the pair likely formed via another mechanism. Furthermore, the separation between both companions at periastron is 43 stellar radii, implying tides are relatively weak and unlikely to change the orbital parameters over billions of years. Main sequence binaries with orbital separation in excess of 10-12 stellar radii show very little signs of tidal evolution post main-sequence [27].

Our photometric monitoring of the system with SPECULOOS [20] and the MONET/South telescope [31] partially covered four eclipse epochs, and have so far failed to recover a second eclipse, a constraint that is included in our orbit fit. We used all of the photometric data to search for evidence of rotationally-modulated variability. While we detect statistically significant 1-2% photometric variability from both sources, we do not detect significant periodicity (see Methods). We determined rotational v sin i values of 8.7±1.0 km/s and 6.8±1.5 km/s for the primary and secondary of 2M1510A from our UVES data, which for our measured radii imply rotation periods of 20-30 hours. This is shorter than the expected convective turnover time for these stars (> 2 days) and argues that both the variability and $H_\alpha$ emission arise from magnetic activity.

The 2M1510 system is a unique, young, substellar triple system, containing a fortuitous equal-mass eclipsing close binary. Another six young brown dwarf triple systems are known (see Supplementary Table 7). Among these, the 2M1510 system stands out in having all three components with approximately the same mass. In addition, the separation ratio between the inner and outer orbits is the smallest of all young triple brown dwarf systems found to date.

The physical parameters and age of 2M1510A are in remarkable agreement with cooling models [2, 3, 4], providing long overdue empirical verification at an age comparable to those of the currently known directly imaged giant exoplanets [32]. Adjustment of the models to explain the luminosity of 2M1510 will also impact the masses and radii of young directly imaged exoplanets, which are inferred from their luminosity and age using similar cooling models to brown dwarfs.

## Acknowledgments

The authors thank the kind personnel of ESO who host SPECULOOS at Paranal Observatory, and who have awarded two DDT programmes to confirm this object (Prog.ID 099.C-0138 and 2100.C-5024, PI Triaud). In addition, we thank Carlos Alvarez, Greg Doppman, Percy Gomez, Heather Hershey, and Julie Rivera at Keck Observatory; and Greg Osterman and Eric Volquardsen at IRTF, for their assistance with the observations reported here. This work also used observations from the LCOGT network, awarded through a DDT programme (PI Alonso).
This research has made use of PyRAF which is a product of the Space Telescope Science Institute, which is operated by AURA for NASA. PyRAF uses IRAF, which is distributed by NOAO, which is operated by AURA, under cooperative agreement with the NSF. We used the SIMBAD database, operated at CDS, Strasbourg, France; NASA's Astrophysics Data System Bibliographic Services; the M, L, T, and Y dwarf compendium housed at DwarfArchives.org; and the SpeX Prism Libraries at http://www.browndwarfs.org/spexprism. The authors recognize and acknowledge the very significant cultural role and reverence that the summit of Mauna Kea has always had within the indigenous Hawaiian community. We are most fortunate and grateful to have the opportunity to conduct observations from this mountain. This research also made use of Astropy (www.astropy.org), a community-developed core PYTHON package for Astronomy as well as the open-source PYTHON packages NUMPY (www.numpy.org), SCIPY (www.scipy.org) and MATPLOTLIB (www.matplotlib.org).

AHMJT has received funding from the European Research Council (ERC) under the European Union's Horizon 2020 research and innovation programme (grant agreement nº 803193/BEBOP). AHMJT also received funding from the Leverhulme Trust under Research Project Grant number RPG-2018-418, and from the Science, Technology and facilities Council grant number ST/S00193X/1. AJB acknowledges funding support from the National Science Foundation under award No. AST-1517177. The material is based upon work supported by the National Aeronautics and Space Administration under Grant No. NNX15AI75G. BOD acknowledges support from the Swiss National Science Foundation (PP00P2-163967). MG received funding from the European Research Council (ERC) under the FP/2007-2013 ERC (grant agreement nº 336480/SPECULOOS), from the ARC grant for Concerted Research Actions, financed by the Wallonia-Brussels Federation, from the Simons Foundation, and from the MERAC foundation. MG and EJ are Senior Research Associates at the F.R.S-FNRS. LD acknowledges support from the Gruber Foundation Fellowship. VKH is supported by a generous Birmingham Doctoral Scholarship and by a studentship from Birmingham's School of Physics & Astronomy.

## Author contribution

AHMJT led the data acquisition, obtained UVES data, and organised the analysis and interpretation of this system. AJB obtained NIRSPEC data, extracted radial-velocities from NIRSPEC and UVES data, produced early parameters of the system, as well as the spectral typing and assessing Argus membership. AB and VKH led the photometric follow-up. AB, ED, CM, PPP, LD and MG reduced the photometric data. VKH produced the global analysis. MG, EJ, DS, BOD, DQ, LD, CM, PPP, JdW, AHMJT, ED, AB, ST participated with the preparation, construction and running of the SPECULOOS facility/survey. JMC provided the DONUTS software used for guiding. DBG provided the SpeX data while RA, FH and TOH participated in the photometric follow-up. VvG calculated stellar models. All authors have assisted with writing the manuscript.

## Competing interest statement

The authors declare not competing financial interests.

## Data availability

All reduced photometric timeseries will be made available for download at the CDS, and on request to the main author. Raw SPECULOOS CCD frames will become available through the ESO archive in January 2021; they can be requested to the authors before this date. Eventually the archive will also contain lightcuves for all reference stars in the frames as well. Our UVES spectra are now publicly available on the ESO archive, and can be found by searching the archive for ProgID 299.C-5046 and 2100.C-5024. MONET-South raw images can be made available upon request. NIRSpec spectra are available at the Keck Observatory Archive, and can be found be searching the archive for PI Burgasser and programs U009, U010 and U136. SpeX spectra are now available via SPLAT (https://github.com/aburgasser/splat).
Requests concerning the data used in this publication can be addressed to Amaury Triaud, and Adam Burgasser.
Figure 1 contains SPECULOOS photometry, as well as UVES and NIRSpec spectra.

## Code availability

The photometric reduction packages use standard public routines such as PyRAF and *astropy*. Radial velocities were extracted from UVES and NIRSpec data from a code built from elements of SPLAT (https://github.com/aburgasser/splat). The *amelie* code combines *ellc* and *emcee* (see text), which are both public codes. *amelie* can be made available upon request, and a stable version is planned to be released on GitHub.

**Figures**

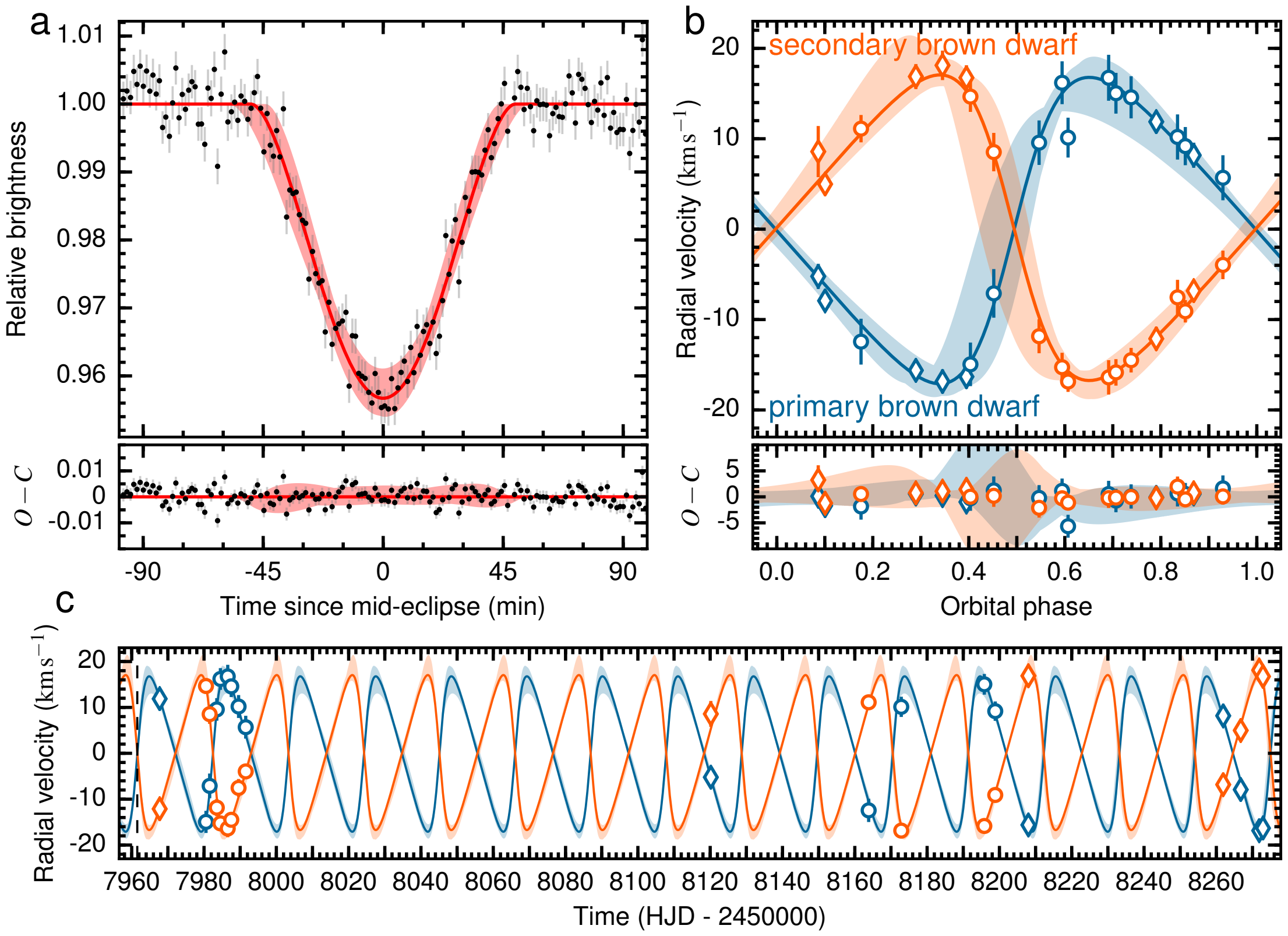

Figure 1. Data demonstrating that 2M1510A is a substellar eclipsing binary. a) SPECULOOS photometric data (black points) and their 1σ uncertainties showing the observed 4% grazing eclipse. The red solid line is the best-fit model from the joint global fit with its 2σ confidence region (red shaded region). b) The phase-folded radial velocity orbit of the brown dwarf primary (blue) and secondary (orange). A systemic velocity of -12.9 km/s has been subtracted from the values. Individual points are measurements (UVES: circles; NIRSPEC: diamonds) with their 1σ uncertainties. Solid lines indicate the joint global fit with 2σ uncertainty regions shaded. c) Radial velocity observations and models as in b), but shown as a function of time. The vertical dash line on day 7961.5 indicates the epoch of the secondary eclipse displayed in a). These combined data allow us to measure the masses of the two brown dwarfs and their combined radius.

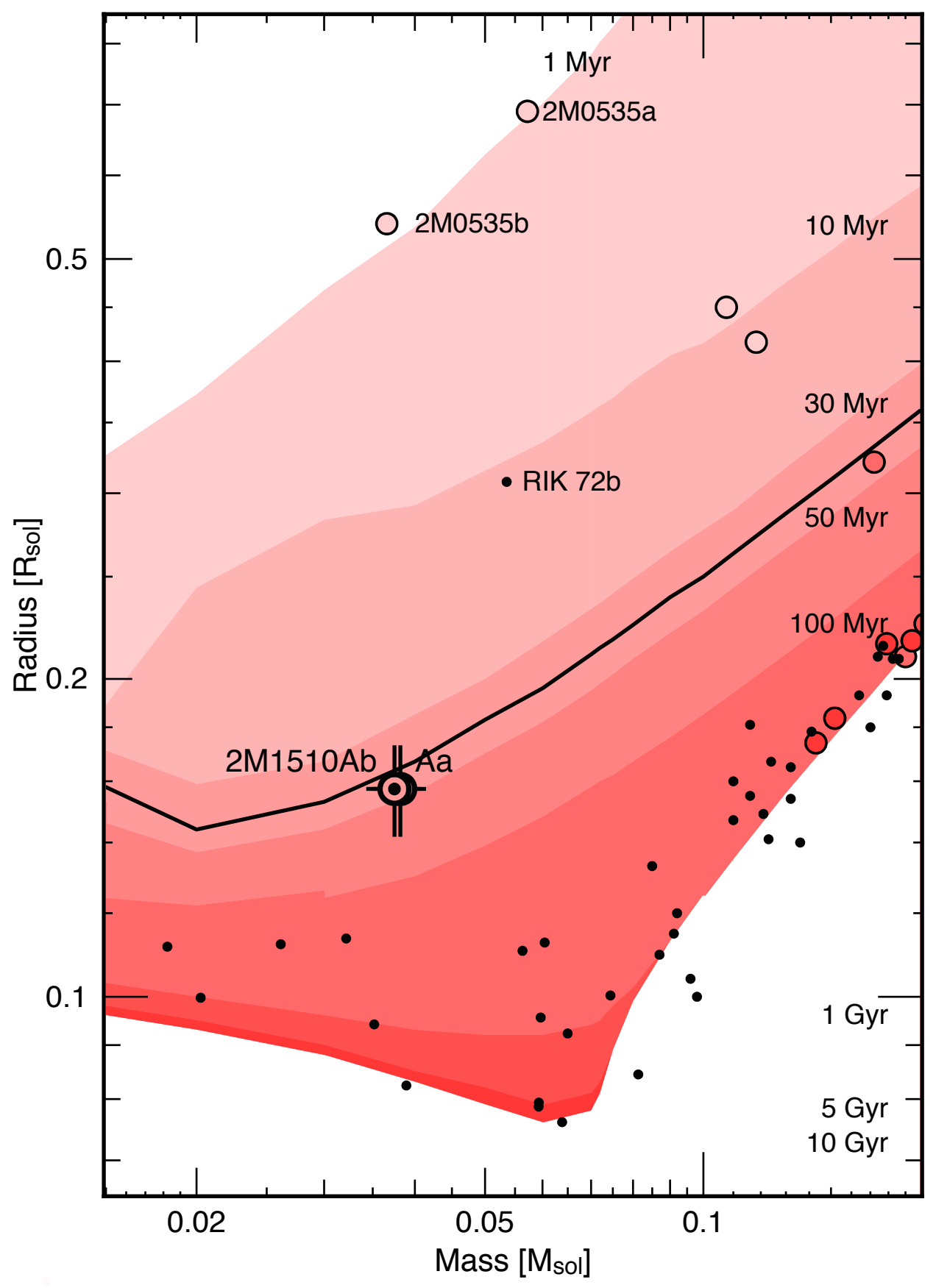

Figure 2. Mass-radius diagram showing the position of the 2M1510Aab eclipsing pair (in black with their 1σ uncertainties). The red shaded areas separate tracks of the Exeter/Lyon evolutionary models [3,4], with ages labelled. Circles delineate mass/radius measurements obtained for previously detected double-line eclipsing binaries with their colour corresponding to the age estimated for each binary. Dark dots depict mass/radius measurements for interferometric measurements, single line eclipsing binaries and transiting exoplanets. The young brown dwarfs 2M0535AB [11] and RIK 27b [7] are explicitly labelled. 2M1510A's parameters are consistent with the 45±5 Myr age of the Argus moving group, just below a 40 Myr isochrone highlighted in black. 2M1510Aab, our double-line eclipsing system is in an area of parameter space with no equivalent systems. Literature measurements are reported in Supplementary Tables 5 and 6.

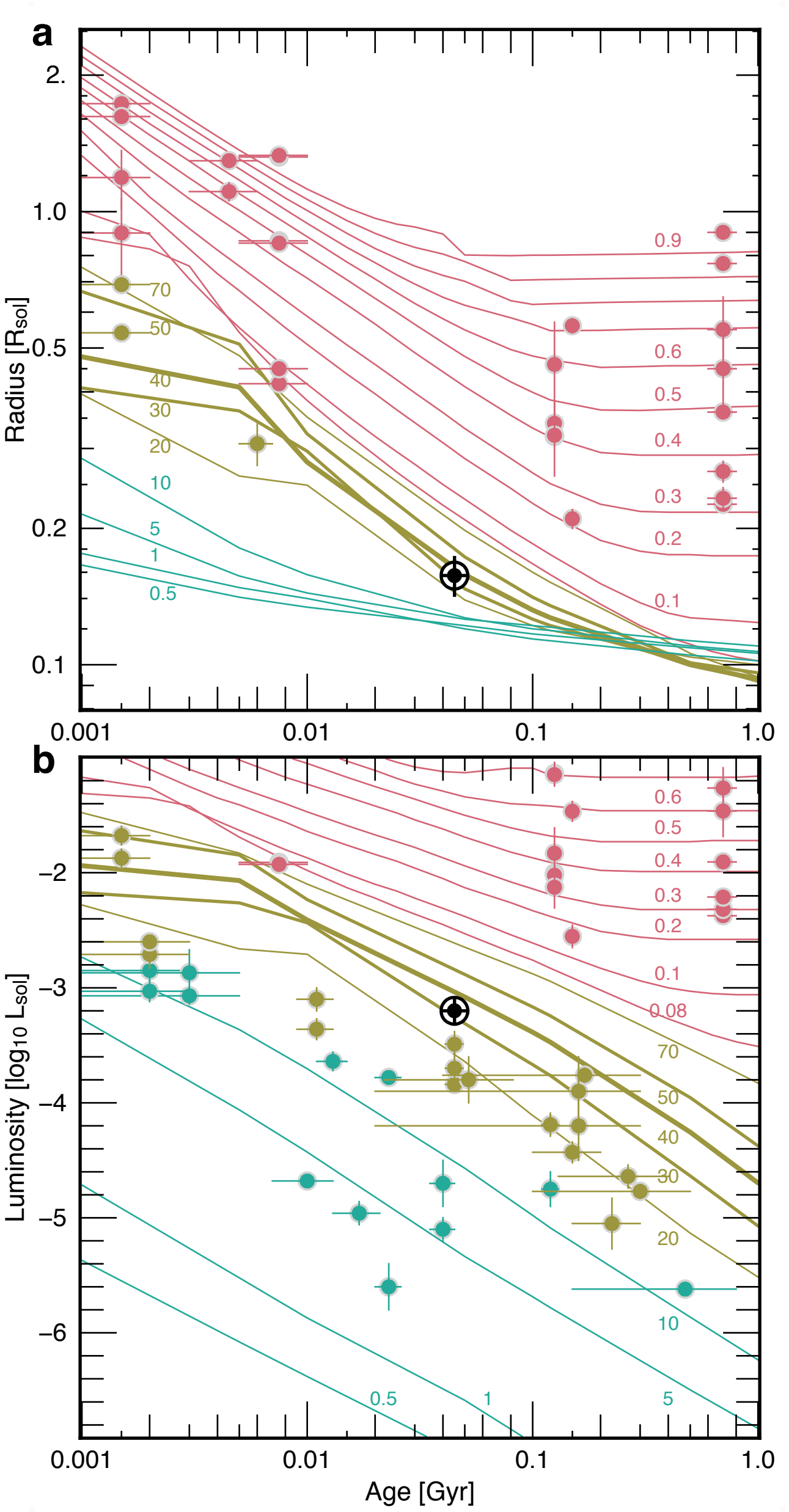

Figure 3. Radius (a) and Luminosity (b) measurements as a function of system age, for stars, brown dwarfs and giant exoplanets with their 1σ uncertainties. Each line corresponds to a model of a given mass [3,4]. Stellar masses (pink lines) are labelled in units of $M_{\odot}$, whereas brown dwarf (khaki lines) and exoplanet masses (blue lines) are labelled in units of $M_{Jup}$. Dots following a similar colour scheme are systems reported in the literature [7,32]. Note that directly imaged planetary mass objects do not have measured radii. 2M1510Aa & Ab are represented as a single black point, and along with 2M0535AB [11] and RIK 72b [7] are the smallest and least massive objects with measured masses, radii and age (panel a). Together they form an empirical isochrone for masses close to 40 $M_{jup}$. Models for 30, 40 and 50 $M_{Jup}$ are depicted as bolder lines. In panel b we note that the bolometric luminosity (and the J, H, K magnitudes) of 2M1510Aab falls under the 0.038 $M_{\odot}$ (40 $M_{Jup}$) track as measured from dynamical masses.

## Table 1

| System information | |
|---|---|
| 2M1510A<br>$15^h10^m47^s.47$ $-28^o18'18.3''$<br><br>2MASS J15104786-2818174[a]<br>*Gaia* 6212595980928732032[b]<br><br>G 17.487 +/- 0.002[b]<br>J 12.84 +/- 0.03[a]<br>H 12.11 +/- 0.03[a]<br>K 11.69 +/- 0.03[a]<br><br>$\pi$ 27.2±0.3 mas[b]<br>$\mu_\alpha \cos\delta$ -118.7±0.5 mas/yr[b]<br>$\mu_\delta$ -46.9±0.4 mas/yr[b]<br>RV -12.9±0.4 km/s<br><br>Spectral Type: M8 (red optical)[c]<br>M9γ (near-infrared) | 2M1510B<br>$15^h10^m47^s.72$ $-28^o18'24.2''$<br><br>2MASS J15104761-2818234[a]<br>*Gaia* 6212595980924278144[b]<br><br>G 18.886 +/- 0.004[b]<br>J 14.01 +/- 0.03[a]<br>H 13.32 +/- 0.03[a]<br>K 12.79 +/- 0.03[a]<br><br>$\pi$ 27.7±0.5 mas[b]<br>$\mu_\alpha \cos\delta$ -117.4±0.9 mas/yr[b]<br>$\mu_\delta$ -45.7±0.7 mas/yr[b]<br>RV -12.0±0.3 km/s<br><br>Spectral Type: M9 (red optical)[c]<br>M9γ (near-infrared) |
| Distance (pc) | 36.6±0.3[b] |
| $\mu_\alpha \cos\delta$ (mas/yr) | -118.4±0.6 |
| $\mu_\delta$ (mas/yr) | -46.6±0.5 |
| RV (km/s) | -12.3±0.4 |
| U, V, W (km/s) | (-5.3±0.3, 24.7±0.3, -1.3±0.2) |
| Age (Myr) | 45±5[d] |
| **2M1510 Aab Orbital and Physical Parameters** | |
| Orbital period (day) | $20.9022_{-0.0056}^{+0.0059}$ |
| Semimajor axis (au) | 0.0627 ± 0.0014 |
| Tsec (HJD-2,450,000) | $7961.53441_{-0.00061}^{+0.00064}$ |
| Eccentricity | 0.309±0.022 |
| Argument of periastron (deg) | -89.9±3.3 |
| Inclination (deg) | 88.5±0.1 |
| Mass ratio | 0.981±0.049 |
| Primary's mass, $M_1$ ($M_\odot$) | $0.0382_{-0.0026}^{+0.0028}$ |
| Secondary's mass, $M_2$ ($M_\odot$) | $0.0375_{-0.0028}^{+0.0029}$ |
| Sum of radii, $R_1 + R_2$ ($R_\odot$) | $0.3147_{-0.0157}^{+0.0159}$ |
| Flux ratio, $f_2 / f_1$ | 0.827±0.013 (819 nm)<br>0.88±0.07 (2.3 μm) |

References: a: [16] b: [8] c: [15] d: [9]

## Methods

### Photometric observations and calibration.

Photometric data reduction of each observing night was performed using the SPECULOOS pipeline based on PyRAF utilities. After standard pre-reduction procedures (bias, dark and flat-field corrections), the images were aligned with respect to the first image of each individual observing run. Aperture photometry was performed on all identified point sources using DAOPHOT [36]. Aperture photometry was done for eight aperture sizes, with sizes dependent on the average full width at half maximum (FWHM) of the stellar profiles through the observing run. The final set of comparison stars and the aperture size were selected to optimize photometric quality, as evaluated by the standard deviation of the flux of a designated check star, i.e. a non-variable star similar in terms of magnitude and colour to the target star. Reduction of data obtained with MONET/South robotic telescope followed a similar procedure, adapted to account for different CCD detector properties.

### Photometric variability of 2M1510.

To study variability over several nights, the data were reduced as will be described in detail in a publication submitted separately (Murray et al.). Briefly, we combine 40 reference stars in the field with similar brightness to build a synthetic timeseries, which is used as a comparison to our main target, 2M1510A. Similarly, we used 132 reference stars for 2M1510B, a larger number due to the fainter magnitude of this source. In addition, we apply a colour-dependent correction to the photometry, based on the precipitable water vapour (PWV) column, a value provided by ESO, at the site where SPECULOOS is located, which allows us to correct for variations in extinction caused by humidity. We further remove 155 frames where the background exceeds 2000 counts, an unusually high value probably caused by light cloud passage. The same procedure is applied for 2M1510A and 2M1510B. Our resulting lightcurves (Supplementary Figure 1) contain 6423 measurements and are stable within 3% over 75 days with no discernible frequency. The only feature of note is the eclipse in the lightcurve of 2M1510A. Over a single night, the variability is typically <1-2%. Larger variations are usually caused by cloudy weather, where our PWV correction is no longer effective.

### Spectral and Gravity Classification of 2M1510A and 2M1510B.

We obtained and analysed new low-resolution near-infrared spectra for 2M1510A and 2M1510B obtained with the SpeX spectrograph on the NASA Infrared Telescope Facility [37]. Prism-dispersed spectra spanning 0.8-2.45 μm at a median resolution $\lambda/\Delta\lambda$ = 120 were obtained on 2015 July 2 (1510A) and 2016 May 28 (1510B) (PI Bardalez Gagliuffi), and reduced using the SpeXtool package [38, 39] following standard procedures. These data were compared to both "field" spectral standards [40] and low-surface gravity M and L dwarf standards [41, 42]. Supplementary Figure 2 shows the best matches, which for both components is the M9 $\gamma$ (very low surface gravity) standard TWA 26 [15]. The spectra of both 2M1510 components are slightly bluer, and have a weaker 1.05 μm VO band, than the spectrum of TWA 26, consistent with an older age than this 10 Myr-old standard [43]. We also evaluated gravity-sensitive spectral indices [41] for both spectra, and find them to be consistent with an intermediate-gravity (INT-G) classification, which equates to an age of order 100 Myr [41, 8]. Thus, both components have spectra consistent with young, low surface gravity brown dwarfs at an age roughly consistent with the Argus association [9].

H$\alpha$ emission for 2M1510A was measured in each of the UVES spectra by first flux calibrating each spectrum to the combined-light absolute *i*-band magnitude of this source, $M_i$ = 14.22±0.02, based on photometry from PAN-STARRS and astrometry from *Gaia* . We used the same magnitude to calibrate each spectrum, and hence ignore any intrinsic variability in the continuum emission. We then directly integrated the H$\alpha$ profile after subtracting the local continuum to compute the total line flux and H$\alpha$ luminosity. We found a mean value of $\log_{10}(L_{H\alpha}/L_{\odot})$ = -7.51±0.19 for the 12 observations (range -7.90±0.27 to -7.37±0.15), with the scatter consistent with a common mean. Using the combined light luminosity of the pair based on the component luminosities computed below, $\log_{10}(2L_{bol}/L_{\odot})$ = -2.90±0.09, we infer $\log_{10}(L_{H\alpha}/L_{bol})$= -4.61±0.20. If the emission arises from just one component, then $\log_{10}(L_{H\alpha}/L_{bol})$ = -4.31±0.20. These values are consistent with prior measurements [43]. The H$\alpha$ line is broader than the velocity difference between the components, with a consistent full-width at 10% maximum of 100±10 km/s, indicative of chromospheric activity. The relative weakness and breadth of the Hα line prevents us from determining which (or both) of the components may be contributing to the emission. If the emission does arise from one component (for example, the secondary), the empirical relations of [12] would predict a $T_{eff}$ decrease of 1-4% and radius increase of up to 6%, corresponding to a flux ratio (0.89-0.97) that roughly reproduces that inferred from the modelled UVES and NIRSPEC observations However, as we cannot constrain the source of the emission, and as the [12] relations are anchored to early- and mid-type field M stars as opposed to late-M young brown dwarfs, we cannot conclusively determine whether this is the source of the observed flux ratio.

**Modelling of VLT/UVES Spectra.**
Thirteen epochs of UVES [21] spectra were obtained between 2017 August 15 and 2018 March 21 (UT) through VLT programs 299.C-5046(A) and 2100.C-5024(A) (PI Triaud). The standard UVES configuration was used with a 0"504 slit and dispersions centred on 437 nm (blue channel) and 760 nm (red channel) with resolutions $\lambda/\Delta\lambda$ = 65,000 (blue channel) and $\lambda/\Delta\lambda$ = 74,000 (red channel). The data were reduced using the ESO UVES pipeline [44]. One of the epochs (2018 February 3) was obtained in poor conditions, and was not included in our analysis; the remaining observations had average signals-to-noise of 6-10 in the red channel and no detection in the blue channel.

The S/N of the UVES data only rises above 5 for λ > 800 nm, and occasionally exceeds 10 in regions centred at 815, 825, 885, 898, 912, 925, and 938 nm. The S/N peaks at these wavelengths due to the blaze pattern of the échelle. We measured radial velocities at the 819 nm Na I doublet which is between the first two “high S/N” peaks. Note that the S/N of each spectrum differs due to variations in observing conditions. In the best spectrum, 30% of the data longward of 800 nm has S/N > 10; in the worst spectrum, none of the data rises above this level.

The extracted spectra were forward-modelled using a Markov Chain Monte Carlo (MCMC) algorithm [47, 48] that is integrated into the SpeX Prism Library Analysis Toolkit (SPLAT) [49]. The modelling was restricted to the 818-820 nm range, which covers the $2p^63p$ ($^2P^0$ J = 1/2, 3/2) -> $2p^63d$ ($^2D$ J = 3/2, 5/2) Na I doublet. Data were initially shifted to the telluric rest frame by cross-correlating superimposed telluric $H_2O$ absorption features in the neighbouring continuum (815-818 nm) to a telluric transmission spectrum $T_0[\lambda]$ [50]. We computed an observed telluric transmission spectrum for each epoch as a function of wavelength λ of the form $T[\lambda] = T_0[\lambda]^\alpha \otimes f_G(\Delta\lambda_{res})$, where α is an exponential scaling factor, ⊗ denotes convolution, and $f_G(v_{res})$ is a Gaussian broadening kernel with width $\Delta\lambda_{res} = \lambda v_{res}/c$; $v_{res}$ was found to be 5.8±0.4 km/s in the continuum region. The telluric model was divided out of the data, with the cores of strong telluric absorption lines (transmission < 70%) masked. Note that in the four 2018 epochs, the telluric lines overlap with the Na I features; however, the breadth of the stellar features still allowed for robust modelling of the data.

The forward model for the data is described as:

$$D[\lambda] = C[\lambda] \times (1+C_{off}) \times \{M_1([\lambda^*_1],T_{eff,1},\log_{10}g_1) \otimes f_V(vsini_1) + f_2 \times M_2([\lambda^*_2],T_{eff,2},\log_{10}g_2) \otimes f_V(vsini_2)\} \otimes f_G(\Delta\lambda_{res}).$$

Here, $M_1$ and $M_2$ are BTSettl08 models [51 - Allard et al. 2012] for the primary and secondary atmospheres with temperatures $T_{eff,1}$ and $T_{eff,2}$ (in K) and log surface gravities $\log_{10}g_1$ and $\log_{10}g_2$ (in cm/s$^2$), respectively; and $0 < f_2 < 1$ is a scaling factor to account for the relative flux of these models. The model wavelength vectors $\lambda^*_1$ and $\lambda^*_2$ are defined as:

$$\lambda^*_1 = \lambda \times (1+[RV_1-RV_{bary}(t)+RV_{shift}]/c)$$
$$\lambda^*_2 = \lambda \times (1+[RV_2-RV_{bary}(t)+RV_{shift}]/c)$$

where $RV_1$ and $RV_2$ are the heliocentric radial velocities of the primary and secondary, $RV_{bary}$ is the barycentric velocity of the system at the observing epoch (t), and $RV_{shift}$ is a nuisance parameter to adjust for zero-point wavelength shifts, all in units of km/s. The model spectra were each convolved with a rotational broadening kernel $f_V(vsini)$ for projected rotational velocities $vsini_1$ and $vsini_2$ (in km/s) for primary and secondary, respectively, using a limb darkening parameter $\varepsilon = 0.6$ [52]. The combined spectrum was convolved with a Gaussian broadening profile $f_G(\Delta\lambda_{res})$ as described above. $C_{off}$ is a nuisance parameter allowing for an additive offset to the flux (e.g., due to residual background emission), while $C[\lambda]$ is a 6th-order polynomial used to match the continuum between model and data.

The full model nominally consists of 20 free parameters; however, the 7 parameters of $C[\lambda]$ were fit at each comparison step, $RV_{bary}$ is pre-determined for each observing epoch, and $\Delta\lambda_{res}$ was adopted from the initial fit of telluric absorption in the neighbouring continuum. Furthermore, to improve convergence we varied the parameters $f_2$, $vsini_1$, and $vsini_2$ only for data from epochs 2017 August 15, 21 and 22, when the velocity separation of the two components was greatest. These values were consistent with each other and yielded means of $f_2 = 0.827\pm0.013$, $vsini_1 = 8.7\pm0.3$ km/s and $vsini_2 = 6.8\pm0.6$ km/s. For the remaining epochs, we forced these three parameters to be fixed. Furthermore, given the near-unity flux ratio, we imposed the constraint $T_{eff,1} = T_{eff,2}$ and $\log_{10}g_1 = \log_{10}g_2$, and constrained 2000 K < $T_{eff,1}$ < 3000 K and 3.5 < $\log g_1$ < 5.5. The remaining 6 free parameters ($T_{eff,1}$, $\log g_1$, $RV_1$, $RV_2$, $RV_{shift}$, $C_{off}$) were determined for each spectrum through forward modelling.

After obtaining initial estimates for the model parameters by visual inspection, we deployed a Metropolis-Hastings MCMC algorithm [53, 54] with Gibbs sampling [55] and a modified Langevin Rule [56] to optimise the fits and explore the parameter space. Due to computational restrictions, we used a single MCMC chain of 3000 steps for each spectrum. At each step, a new value for a single parameter was proposed by drawing from a Gaussian distribution centred on the current parameter value and with a fixed (pre-determined) distribution width. We used the F-test cumulative distribution function $F_{CDF}(\chi^2_p/\chi^2_c, DOF, DOF)$ as our test condition, comparing the $\chi^2$ values between the current ($\chi^2_c$) and proposed ($\chi^2_p$) model with degrees of freedom DOF = 750. The DOF is equal to the number of spectral data points between 818-820 nm (depending on telluric absorption masking) minus the number of fit parameters, which includes the 7 polynomial coefficients of $C[\lambda]$. The proposed parameter set was accepted if $2F_{CDF} - 1 < U(0,1)$, where U(0,1) is a random number drawn from a uniform distribution between 0 and 1. With this criterion, every model with smaller $\chi^2$ was accepted, while higher $\chi^2$ models were accepted based on the degree of deviation from the current model as measured by $F_{CDF}$. We added an additional criterion to check if the chain was wandering significantly from the current best-fit solution by calculating $F'_{CDF}(\chi^2_c/\chi^2_{min}$, DOF, DOF) and requiring that $F'_{CDF} < 0.95$; i.e., that the chain is no more than 95% ($2\sigma$) away from the best-fit model. If this condition was violated, the chain was forced to return to the best-fit parameter set. This algorithm had a high acceptance ratio of 90%, implying that it efficiently explored the parameter space. Tests of shorter chains, as well as the final overall parameter distribution, indicate successful convergence for each parameter. The final parameters were determined as the median and 16% and 84% quantiles of the chain after removing the initial 25% of samples to allow for convergence.

Supplementary Figure 3 & 4 illustrate the best-fit spectrum and parameter distributions, respectively, for the spectrum obtained on 2017 August 21 (UT). A table summarizing the parameters inferred from each of the UVES epochs is given in Supplementary Table 1. In all fits, the best-fit $\chi^2$ is consistent with accurate representation of the data given the fit DOFs, while global model variables $T_{eff,1}$ = 2753±196 K and $\log_{10}g_1$ = 4.10±0.24 are consistent between all 12 epochs. Note that individual epoch uncertainties in these parameters were increased to reflect the scatter in the measurements. There are no strong correlations in the posterior distributions between the inferred atmospheric parameters ($T_{eff,1}$ and $\log_{10}g_1$) and component radial velocities, so these parameters are effectively decoupled.

**Acquisition and Modelling of Keck/NIRSPEC Spectra.**
Seven epochs of observations of 2M1510A and three epochs of 2M1510B were obtained with Keck/NIRSPEC [20] between 2017 August 2 and 2018 June 3 (UT; PI Burgasser). The observations were obtained using NIRSPEC's high-resolution mode and N7 filter with either the 0"432 wide slit (2017 August 2) or 0"288 wide slit (all other epochs), providing coverage over 1.99-2.40 μm in 7 orders with average resolutions $\lambda/\Delta\lambda$ = 25,000 (0"432 slit) and $\lambda/\Delta\lambda$ = 37,500 (0"288 slit). Spectral data were extracted using a custom modification of the REDSPEC code [57], which includes spatial and spectral rectification, optimal extraction, and combination of AB pairs. Nearby A0 V stars were observed before or after 2M1510A and 2M1510B for telluric calibration and wavelength calibration (see below).

Our analysis focused on the 2.29-2.32 μm spectral region which encompasses both strong CO absorption bands intrinsic to the stars and telluric $H_2O$ and CO features used to refine the wavelength calibration. The data for 2M1510A were forward modelled in the same manner as the UVES data, with the exception that $H_2O$ and CO telluric absorption features were left unmasked in the data, and the forward model included the telluric transmission function. The forward model took the form:

$$D[\lambda] = C[\lambda] \times (1+C_{off}) \times \{T_0[\lambda]^{\alpha} \times$$
$$[M_1([\lambda^*_1],T_{eff,1},\log_{10}g_1) \otimes f_V(\mathrm{vsini}_1) + f_2 \times M_2([\lambda^*_2],T_{eff,1},\log_{10}g_1) \otimes f_V(\mathrm{vsini}_2) ]\} \otimes f_G(\Delta\lambda_{res})$$

with parameters as described above. Since the data are too low resolution to measure vsini for either component, we fixed these parameters to the best-fit values from the UVES data, and again assumed common temperatures and surface gravities. Additionally, we only fit $f_2$ for data with the maximum velocity separations, on 2017 August 2, 2018 March 30 and 2018 June 2 (UT). This resulted in 8 or 9 free parameters ($T_{eff,1}$, $\log_{10}g_1$, $f_2$, $RV_1$, $RV_2$, $\alpha$, $RV_{shift}$, $C_{off}$, $\Delta\lambda_{res}$) for each spectrum that were determined through MCMC forward modelling. As with the UVES data, we used a single MCMC chain with the evaluation metrics described above.

Supplementary Figures 5 and 6 illustrate the best-fit spectrum and parameter distributions, respectively, for the spectrum of 2M1510A obtained on 2017 August 2 (UT), and Supplementary Table 2 summarizes the parameter determinations for each NIRSPEC epoch. Again, the best-fit $\chi^2$ values were consistent with this model being an accurate representation of the data, and we found mean values of $T_{eff,1}$ = 2755±20 K, $\log_{10}g_1$ = 5.17±0.12 and $f_2$ = 0.88±0.07 to be consistent between the seven epochs. Both $T_{eff,1}$ and $f_2$ are consistent with the UVES observations, but $\log_{10}g_1$ is significantly higher, suggesting that the models may be gravity-insensitive in this wavelength regime. We found no significant correlations between surface gravity and component radial velocities in our analysis.

For 2M1510B, we applied the same analysis using a single component model, which took the form:

$$D[\lambda] = C[\lambda] \times (1+C_{off}) \times \{ T_0[\lambda]^{\alpha} \times M([\lambda^*],T_{eff},\log_{10}g) \otimes f_V(vsini) \} \otimes f_G(\Delta\lambda_{res}).$$

In this case, we left vsini as a free parameter, resulting in 8 free parameters ($T_{eff}$, $\log_{10}g$, RV, vsini, α, $RV_{shift}$, $C_{off,}$, $\Delta\lambda_{res}$) determined through MCMC forward modelling. Supplementary Figures 7 and 8 illustrate the best-fit spectrum and parameter distributions, respectively, for the spectrum of 2M1510B obtained on 2017 August 2 (UT), and Supplementary Table 3 summarizes the parameter determinations for each NIRSPEC epoch and mean values. The average radial velocity inferred for 2M1510B, -12.0±0.3 km/s, is within 2σ of the systemic velocity of 2M1510A (see below).

**Membership in the Argus Association.**

Given the low surface gravity features present in the near-infrared spectra of 2M1510A and 2M1510B, and prior analysis indicating membership of 2M1510A in Argus [8], we re-evaluated the spatial and kinematic alignment of these two sources with 29 nearby young associations using the Banyan Σ code [58], updated to account for a new Argus membership analysis [9]. We adopted as systemic values the uncertainty-weighted averages of the radial velocities, and *Gaia* parallaxes and proper motions of the two components (Table 1). This analysis confirmed membership in the Argus Association with 89% probability, versus 11% probability of being a field star contaminant. The roughly 40 members of Argus have a mean distance of 72 pc from the Sun, and a consensus isochronal age of 40-50 Myr [70, 71, 58; 19]. Hence, membership is consistent with the surface gravity classifications of the low-resolution spectra of 2M1510A and B, which are intermediate between VL-G (≈ 10 Myr) and INT-G (≈100 Myr; [41]); as well as the age inferred from the evolutionary models based on our mass and radius constraints (Figure 2).

**Global analysis of the radial velocity and photometric data.**

The orbital and physical parameters of the 2M1510A binary were extracted by performing a global analysis of the radial velocities and photometry. We assumed the motion of the stars followed Keplerian orbits, and considered both the Doppler variation in their radial motions and geometric obscuration during eclipses phases. Posterior probability distributions of the system's parameters were obtained using a global Markov Chain Monte Carlo (MCMC). The binary lightcurve software *ellc* was used to generate orbital and light curve models [22], and we employed the affine-invariant MCMC sampler *emcee* to explore parameter space [23]. We validated our approach [60, 14] by re-analysing data from the eclipsing binary system WW Aurigae [61] and found a 1σ agreement with reported values.

The MCMC's fit parameters were the orbital period, P; the time of secondary eclipse, $T_{sec}$; the sum of scaled radii, $r_1 + r_2 = (R_1 + R_2)/a$, where $R_1$ is the primary radius (defined as the most massive of the pair), $R_2$ is the secondary radius, and a is the semimajor axis; the cosine of the orbital inclination, cos i; the eccentricity parameters $\sqrt{e} \sin \omega$ and $\sqrt{e} \cos \omega$, where ω is the argument of periapsis; and the radial velocity semi-amplitudes $K_1$ and $K_2$ for the primary and secondary, respectively. In addition, we used a linear regression at each MCMC step to calculate a normalisation factor for each individual lightcurve, and a velocity offset for each component to marginalise over any difference. We also allow an offset for each spectrograph. We did not apply any specific prior to any of our parameters in our initial runs, but identified several simplifications with minimal impact on the results (see next subsection).

An initial MCMC analysis of only the radial velocities provided initial estimates of the orbital period and non-zero eccentricity, and independently confirmed that the observed eclipse happened at superior conjunction near periastron. With these initial constraints, our full MCMC employed 200 walkers of 20,000 steps each, providing a good exploration of parameter space. The convergence of the chain was assessed using the Gelman-Rubin statistic [62]. All our parameters reached the recommended R-hat < 1.1, which is considered converged [63]. We thinned our chains by a factor of 200 due to autocorrelation and were left with 9000 independent samples for each parameter. Quantiles at 16%, 50% and 84% are reported in Supplementary Table 4. We graphically represent the best solution and its 2σ confidence region in Figure 1. We converted our fit parameters into physical parameters, and report them in Table 1 and Supplementary Table 4.

**Assumptions taken for the final orbit fit.**

We fixed a number of parameters in our final MCMC, after testing that they had no effect on the fit. As suspected from the grazing eclipse, all runs converged to an orbital configuration where the combination of low inclination and high eccentricity implied the absence of a visible primary eclipse. This means we cannot measure the ratio of eclipse depths, and hence the individual radii.

The surface brightness ratio, radius ratio, and limb darkening coefficients of the brown dwarfs remained unconstrained, resulting in flat posterior distributions. We verified this by comparing the Bayesian Information Criterion (BIC [64]), finding no instances where any combination of these parameters was preferred over fixing them ($\Delta$BIC < 6). Consequently, we fixed the brightness ratio $f_2/f_1 = 0.8$, based on the spectroscopic flux ratios obtained from the UVES and NIRSPEC observations. The radius ratio is fixed to $R_2/R_1 = 1$ since the mass ratio is also compatible with unity. Further support for this assumption emerges from the Exeter/Lyon models [4], where brown dwarfs of the age of Argus and within the mass ranges we infer, share the same size (see Fig. 3a). We adopt a quadratic limb darkening law and determine its limb darkening coefficients c1 and c2 with LDTk [65]. These coefficients were calculated using a stellar spectrum library [66], integrating over a red-optical/near-infrared bandpass (I+z) attenuated by the quantum efficiency of SPECULOOS's e2V deep-depleted detector, assuming evolutionary model parameters for our 45 Myr brown dwarf system: $T_{eff} = 2400$ K, $\log_{10} g = 4.55$, and [M/H] = 0.0. Although the stellar parameters may deviate slightly from these values, the limb darkening coefficients are relatively insensitive over a wide range of temperature and surface gravity.

Our fit to the photometric data did not include detrending of the lightcurves. We attempted to detrend the photometric data by constructing a baseline model of time, sky background levels, FWHM changes in the PSF, and changes in pixel position x,y, using polynomials of degrees 1–3 for combinations of these parameters. For each baseline model we calculated the BIC, in which the lowest value would be preferred. For all data used in our analysis, we found that no photometric detrending was justified. We allow the fit to rescale our photometric and radial velocity uncertainties.

**Bolometric Luminosity and Absolute JHK magnitudes.**

Absolute magnitudes (M) for 2M1510A were computed from apparent magnitudes (m) and the *Gaia* parallax distance (d) using the standard equation $m - M = 5 \log_{10}(d/10\ \mathrm{pc})$. For 2MASS photometry, we infer combined-light absolute magnitudes of $M_J = 10.02 \pm 0.05$, $M_H = 9.29 \pm 0.04$, and $M_K = 8.87 \pm 0.04$. Assuming both components have approximately the same near-infrared brightness, corresponding component absolute magnitudes are $M_J = 10.78 \pm 0.05$, $M_H = 10.05 \pm 0.04$, and $M_K = 9.63 \pm 0.04$.

To calculate the bolometric luminosity, we used the J-band bolometric correction relations of [67], assuming a spectral type of M8.5±0.5 (the average of VL-G and FLD-G classifications). Using $BC_J = 1.97 \pm 0.16$ and the component absolute J-band magnitude given above, we find $\log_{10}(L_{bol}/L_\odot) = -3.20 \pm 0.07$, again assuming equal-brightness components. To account for a K-band flux ratio of 0.88, we raise the uncertainties to reach the value 3.20±0.09.

Assuming a mass of 0.04 $M_\odot$ for both components, evolutionary models [4] predict $M_J = 10.13 \pm 0.12$, $M_H = 9.57 \pm 0.11$, and $M_K = 9.23 \pm 0.11$ at the 45±5 Myr age of Argus. The Observed-Expected differences are thus $\Delta M_J = 0.65 \pm 0.14$, $\Delta M_H = 0.48 \pm 0.13$, and $\Delta M_K = 0.40 \pm 0.13$, all significantly fainter. The ΔM values were estimated assuming both objects have the same luminosity. However our spectroscopic fits indicate a flux ratio $f_2/f_1$ = 0.83-0.88. This translates into a 0.07-0.10 magnitude change in the absolute magnitudes, which are smaller than the current uncertainties. If we use the absolute magnitudes to infer mass, the evolutionary models predict 0.025-0.030 $M_\odot$ for each component, 20% to 35% lower than the dynamical masses. Our estimate of bolometric luminosity (based on K magnitudes) leads to photometric masses of $0.31^{+0.04}_{-0.03}$ $M_\odot$ (Figure 2). If instead, we assume the bolometric luminosity and dynamical masses are correct, the models would imply that the system has an age of $71^{+15}_{-12}$ Myr, consistent with an older (but not consensus) age for Argus of 60 Myr [41]. However, the observed sum of radii would now be larger by a factor 1.09 compared to the models.

**Kozai analysis.**

It has been posited that most binaries with orbital periods below 10 days were formed via the Lidov-Kozai mechanism [29, 30]. If inclined by more than 40º to the inner binary, a tertiary can exchange angular momentum with the inner pair. The eccentricity of the inner pair oscillates, and is able to reach values near unity under some circumstances. When the two inner stars are at periastron, tidal friction will circularise the orbit and shrink its semi-major axis. Currently 2M1510Aa and 2M1510Ab are far enough from each another at periastron that tidal forces are irrelevant. Both objects are separated by 43 radii at periastron and tides are mostly ineffective for separations > 10-12 radii, as tidal circularisation timescales are proportional to $R^5$ [68]. However, the system is young enough that it could currently be undergoing Lidov-Kozai oscillations depending on the orbital elements of the tertiary companion.

We estimate the Lidov-Kozai oscillation timescale [69] for the triple system, defined as

$$\tau_{LK} = (2\,\pi/3) * (P_{out}^2 / P_{in}) * ((M_1 + M_2 + M_3)/M_2) * (1-e_{out}^2)^{3/2}$$

With $P_{out}$ and $P_{in}$ the outer and inner orbital periods, respectively; $e_{out}$, the outer eccentricity; and $M_1$ and $M_2$ the masses of 2M1510Aa & Ab, and $M_3$ the mass of 2M1510B. We take $M_1$, $M_2$, and $P_{in}$ from Table 1, assume $M_3 = M_1$, and assume $e_{out} = 0$. Using Kepler's third law and the projected separation of 2M1510A and 2M1510B (250 AU) as the tertiary semi-major axis, we calculate $P_{out}$ ~ 11,500 years. For these values, the Lidov-Kozai timescale is of order 1 Gyr, implying that the outer tertiary is unlikely to be responsible for the elevated eccentricity of the inner pair. Exploring our assumptions (Supplementary Figure 7), we find that only for $e_{out} > 0.8$ and 2M1510B currently near apastron would $\tau_{LK}$ approach the age of the system. Given the narrow parameter space of this solution, we conclude that the tertiary had no influence on the eccentricity of 2M1510Aa and 2M1510Ab, and that this eccentricity is likely primordial.

In addition, in the absence of a credible Lidov-Kozai path, the 2M1510 pair must have come together via an alternative process, for instance disc fragmentation and a subsequent disc migration [72].

## Methods References

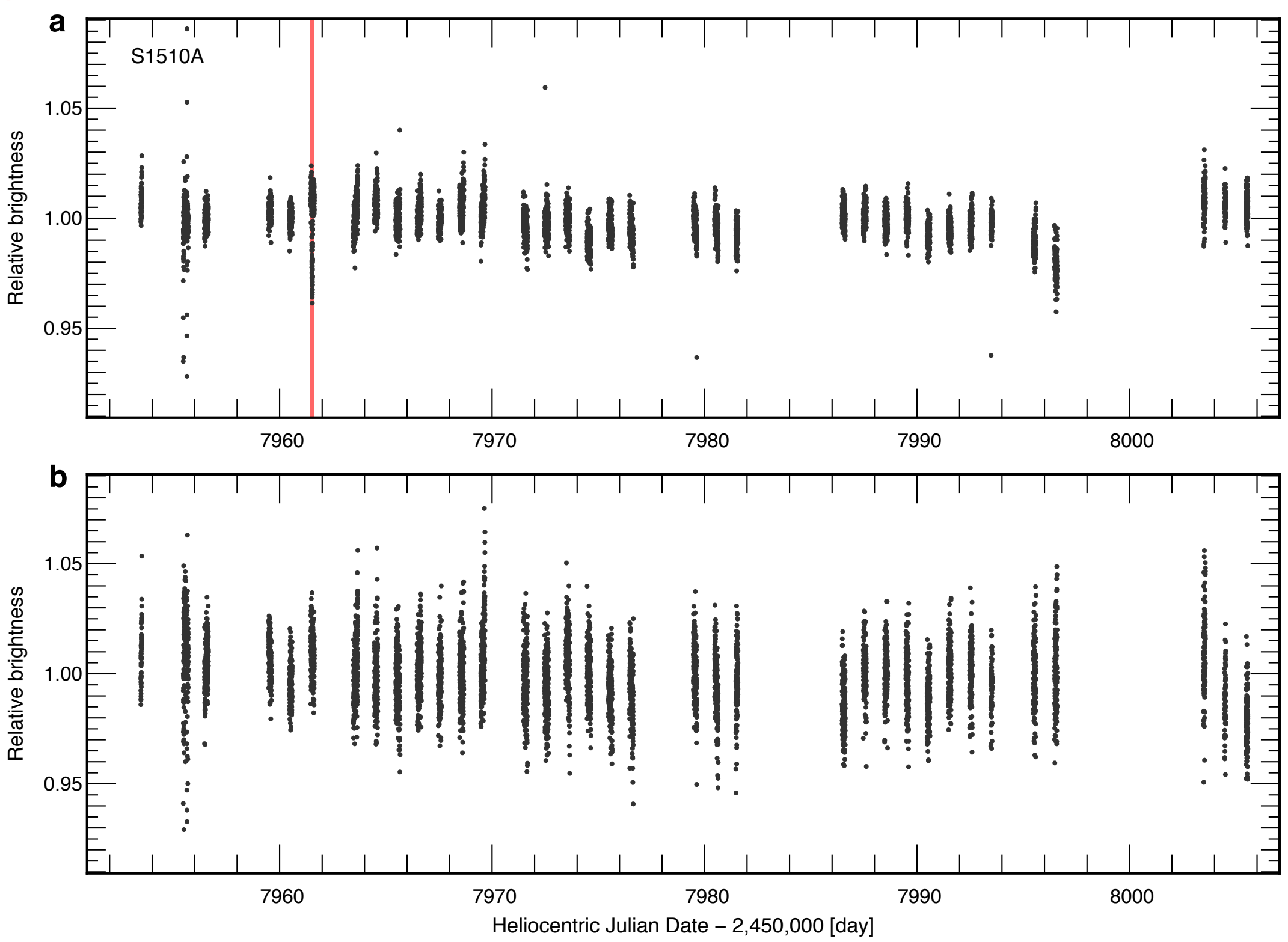

Supplementary Figure 1: Relative flux as a function of time for 2M1510A (a), and 2M1510B (b) showing the variability of the sources on a subset of our lightcurve. No dominant period is detected. The red area indicates the epoch of the observed secondary eclipse.

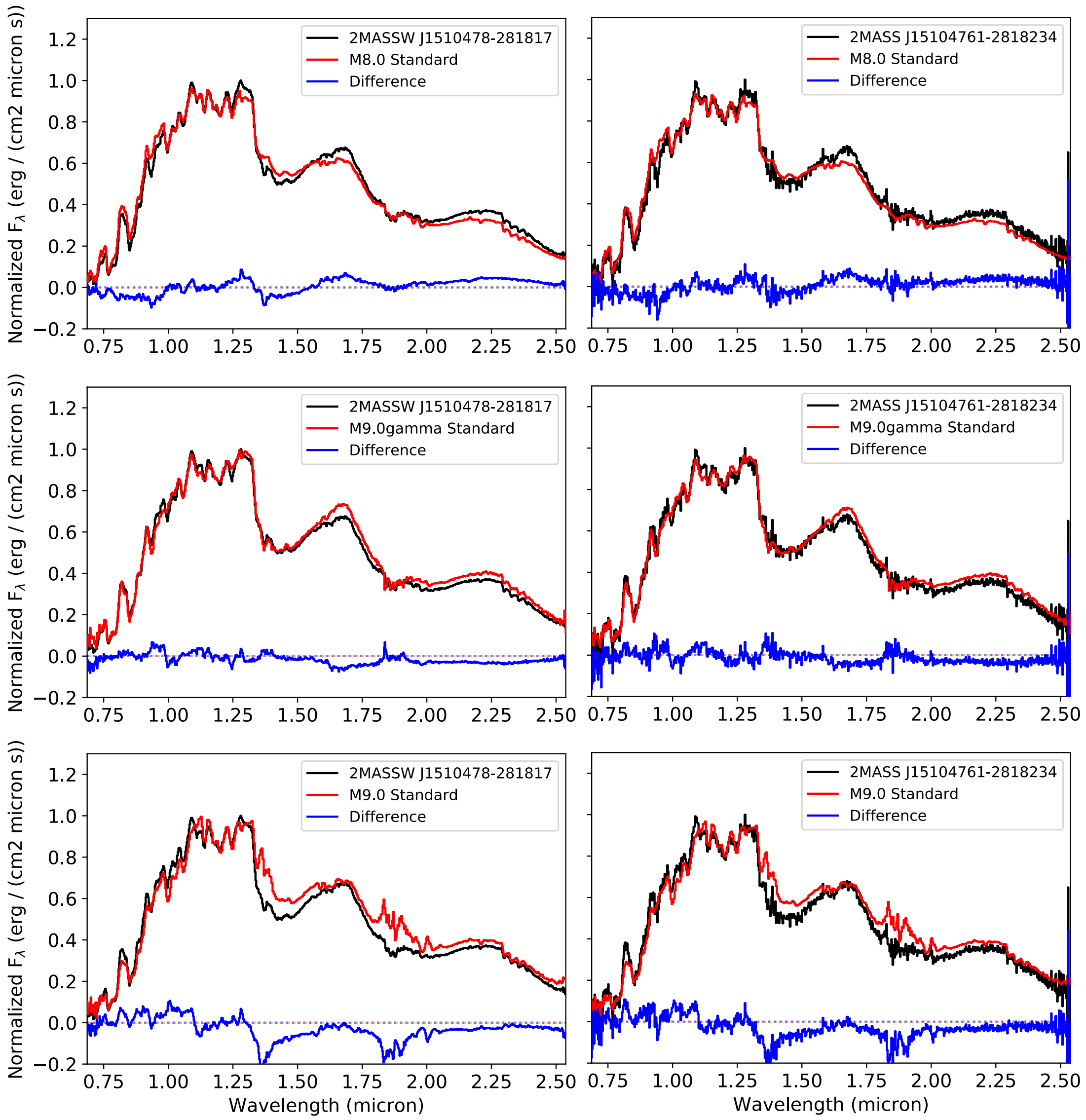

Supplementary Figure 2. Normalized, low-resolution near-infrared spectra (black lines) of 2M1510A (left) and 2M1510B (right), compared to three spectral standard. The M9 $\gamma$ TWA 26 (red lines [44 - Looper et al. 2007]) is superior to both best-fit field and intermediate-gravity spectral standards for both sources, confirming the youth of the two components.

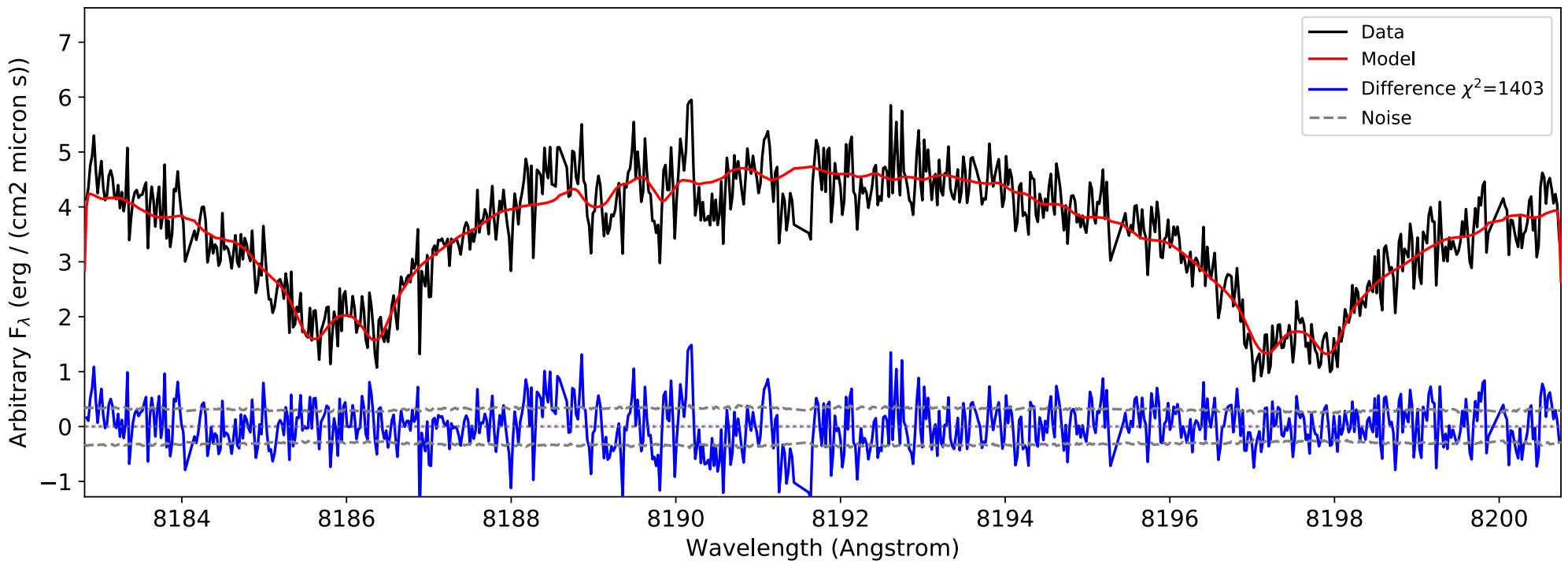

Supplementary Figure 3. Best forward-model fit (red line) to the UVES spectrum of 2M1510A obtained on 2017 August 15 (black line). The difference spectrum (blue line) is consistent with the noise (grey dashed lines ±1σ). Regions of strong telluric absorption have been masked.

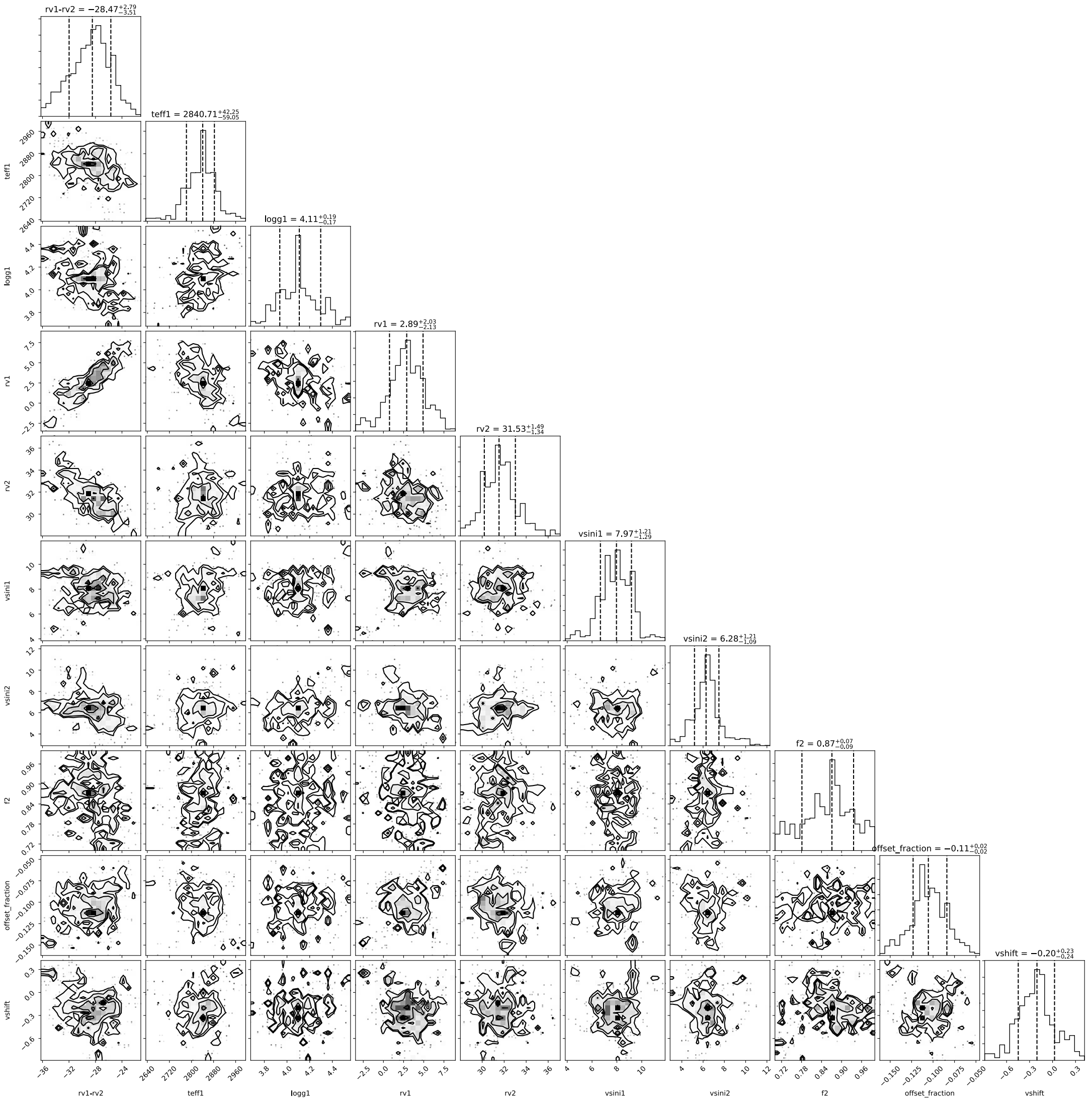

Supplementary Figure 4. Corner plot showing the parameter posterior distribution of our MCMC analysis of the UVES spectrum of 2M1510A obtained on 2017 August 15 (UT). The panels along the right diagonal display the marginalized distributions of each parameter, with median and 16% and 84% quantile uncertainties labeled. The inner panels show two-parameter correlations with all other parameters marginalized. The parameters shown are the stellar model effective temperature (teff1) and surface gravity (logg1), primary (rv1) and secondary (rv2) radial velocities, primary (vsini1) and secondary (vsini2) rotational velocities, secondary flux ratio (f2), fractional additive offset for the model (offset_fraction = $C_{off}$), and overall spectrum velocity shift (vshift = $RV_{shift}$). The radial velocities shown do not take into account the barycentric motion at this epoch (-28.944 km/s).

## Supplementary Table 1: Parameter values for fits to UVES observations of 2M1510A

| HJD - 2,400,000 | $T_{eff,1}$ (K) | $log_{10}g_1$ (cgs) | $RV_1$ (km/s) | $RV_2$ (km/s) | $f_2$ | $vsini_1$ (km/s) | $vsini_2$ (km/s) |
|---|---|---|---|---|---|---|---|
| 57980.54235 | 2841+42-59 | 4.11+0.19-0.17 | -26.1+2.0-2.1 | 2.6+1.5-1.3 | 0.87+0.07-0.09 | 7.9±1.3 | 6.4±1.4 |
| 57981.55828 | 2614+53-50 | 3.99+0.21-0.15 | -20.7+2.0-2.0 | -4.5+2.0-1.6 | | | |
| 57983.53136 | 2580+37-55 | 3.92+0.19-0.16 | -3.0+1.9-1.9 | -24.7+1.6-2.1 | | | |
| 57984.53233 | 2897+79-71 | 4.33+0.21-0.22 | 4.0+2.2-1.9 | -27.6+2.2-1.6 | | | |
| 57986.55909 | 2822+50-69 | 4.17+0.25-0.19 | 3.6+2.1-2.0 | -28.7+1.8-2.1 | 0.79+0.10-0.06 | 9.5±1.2 | 5.7±1.6 |
| 57987.53693 | 2790+38-86 | 4.00+0.22-0.18 | 0.3+1.8-1.6 | -28.1+1.8-1.9 | 0.80+0.11-0.07 | 8.7±1.5 | 8.8±1.8 |
| 57989.55772 | 2598+70-71 | 3.99+0.28-0.17 | -3.2+2.2-2.7 | -19.7+1.9-1.9 | | | |
| 57991.53363 | 2857+68-53 | 4.12+0.18-0.14 | -7.3+1.7-2.2 | -16.4+2.2-1.9 | | | |
| 58163.83375 | 2456+69-94 | 3.82+0.15-0.13 | -26.2+1.6-2.1 | -3.4+1.4-2.2 | | | |
| 58172.84604 | 2817+74-98 | 4.13+0.23-0.19 | 3.4+1.3-1.2 | -29.0+1.4-1.3 | | | |
| 58195.82493 | 2974+20-55 | 4.27+0.10-0.10 | 2.7+1.6-1.5 | -28.9+1.6-1.5 | | | |
| 58198.84487 | 2758+99-134 | 4.06+0.22-0.08 | -5.2+1.4-1.1 | -23.1+1.5-1.5 | | | |
| MEAN | 2753±196 | 4.10±0.24 | | | 0.827±0.013 | 8.7±1.0 | 6.8±1.5 |

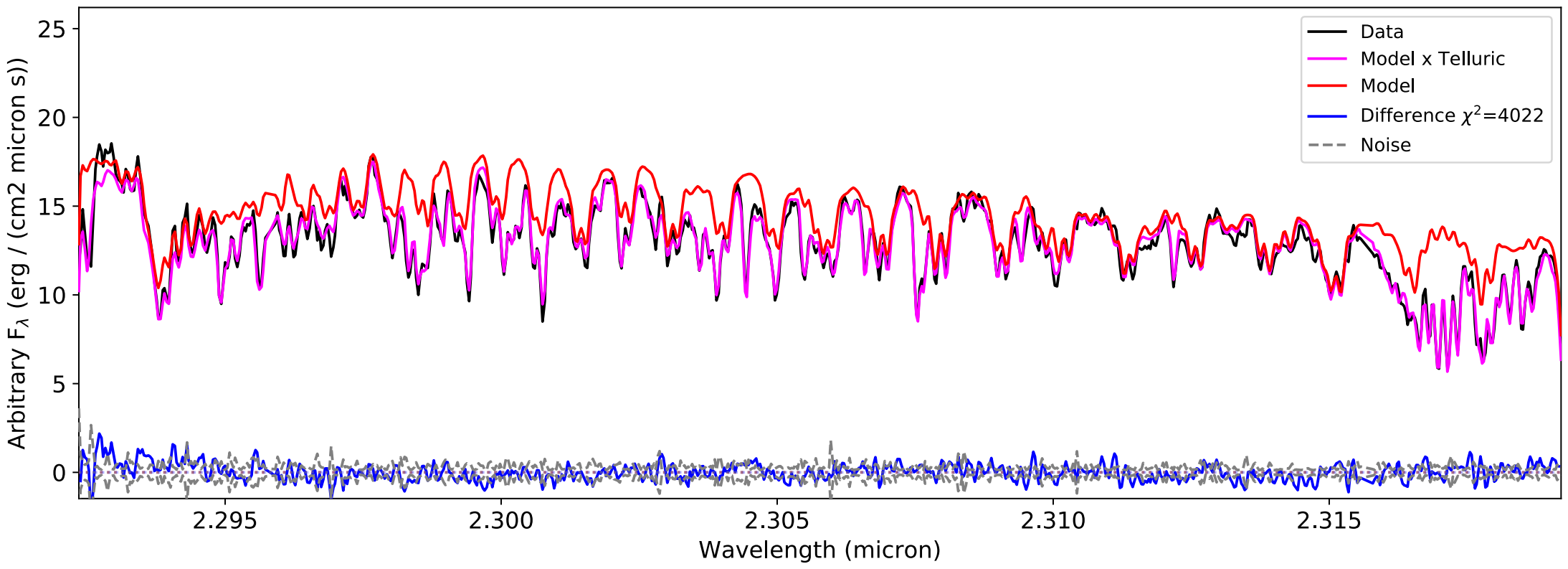

Supplementary Figure 5. Best forward-model fit (red line without telluric absorption; magenta line with telluric absorption) to the NIRSPEC spectrum of 2M1510A obtained on 2017 August 2 (UT; black line). The difference spectrum (blue line) is consistent with the noise (grey dashed lines ±1σ).

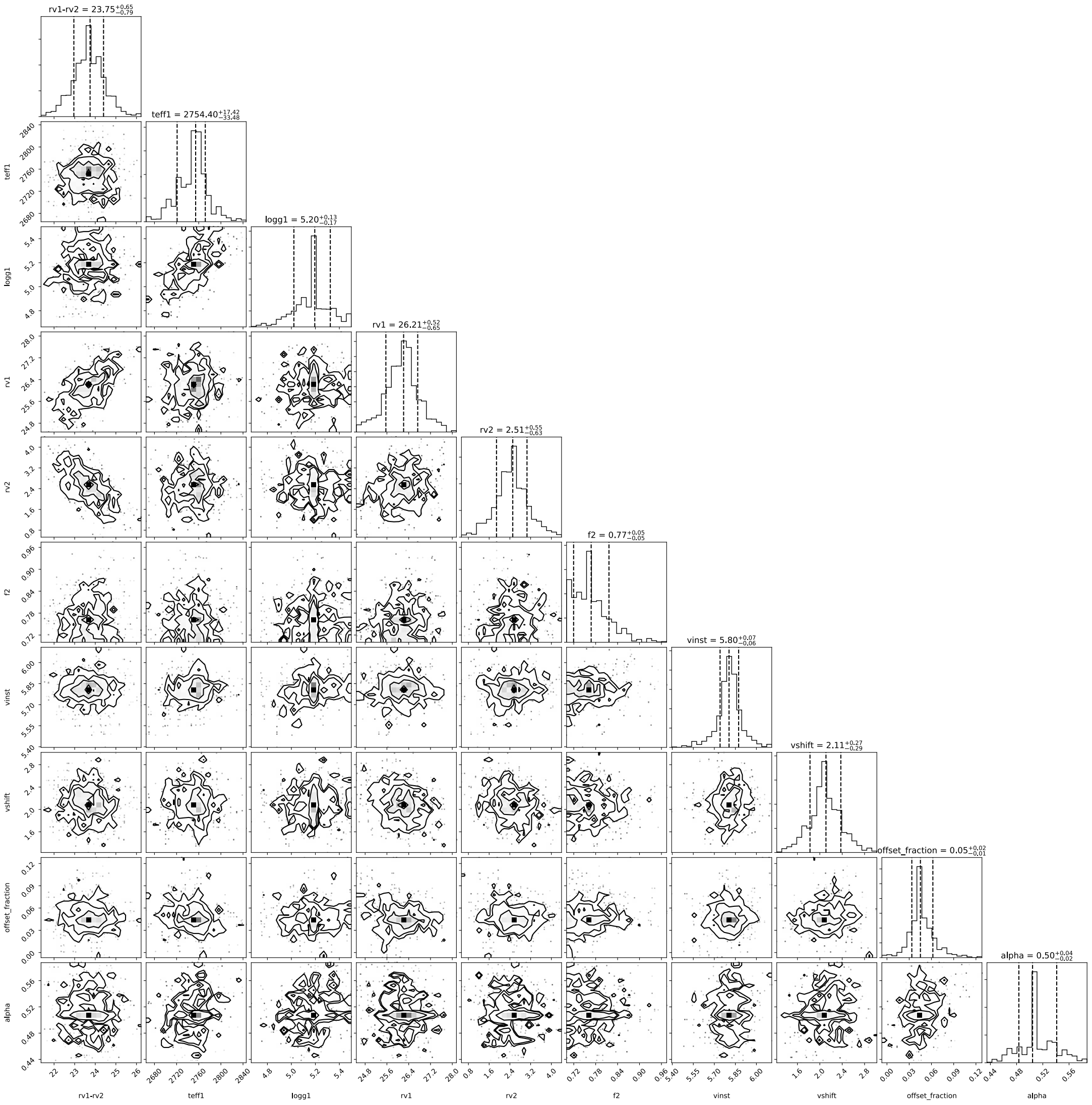

Supplementary Figure 6. Same as Supplementary Figure 4 for the MCMC analysis of the NIRSPEC spectrum of 2M1510A obtained on 2017 August 2 (UT). The radial velocities shown do not take into account the barycentric motion at this epoch (-23-131 km/s).

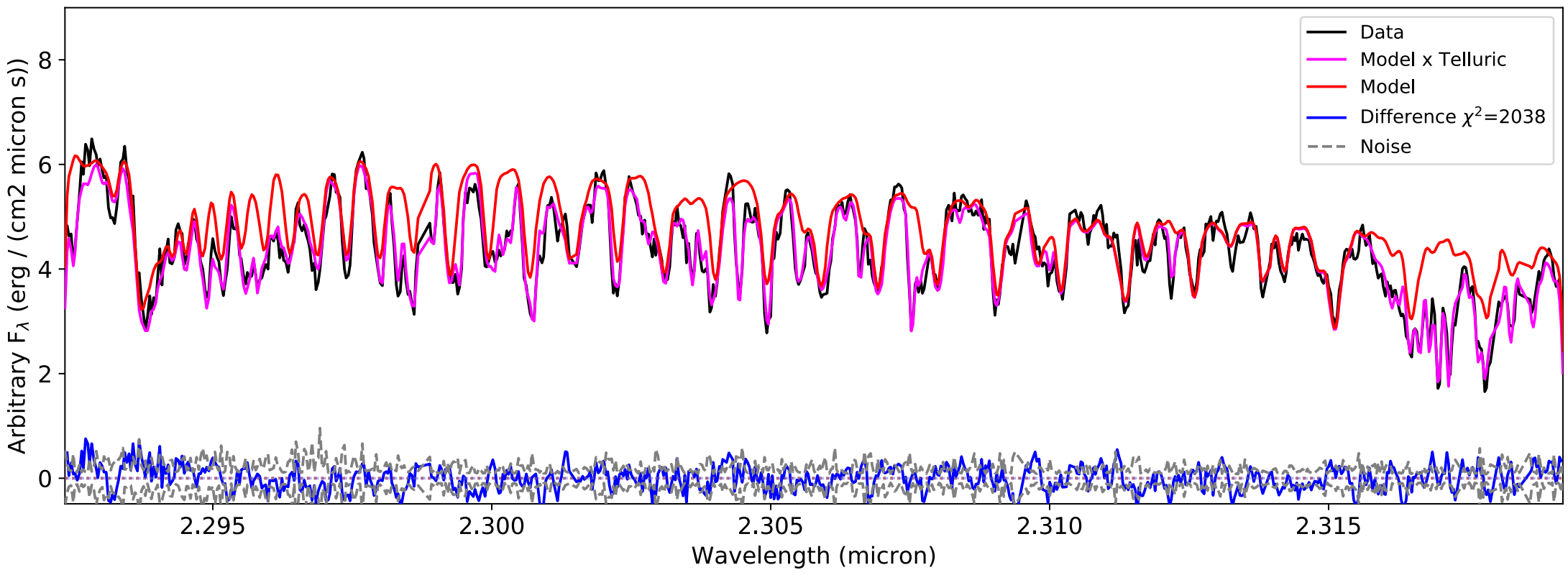

Supplementary Figure 7. Same as Supplementary Figure 5 for spectral data of 2M1510B, in this case fit to a single component model.

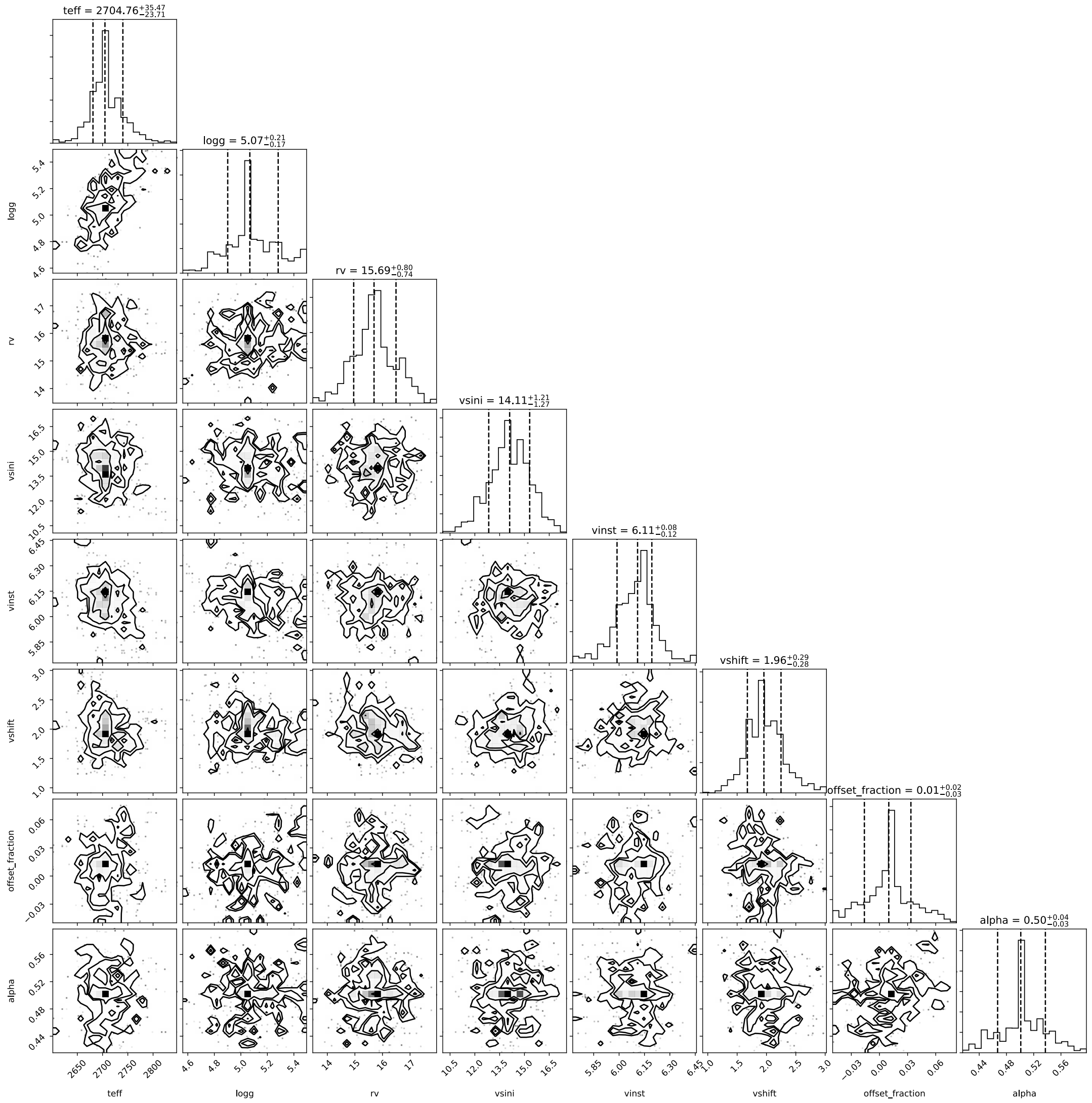

Supplementary Figure 8. Same as Supplementary Figure 6 for the MCMC analysis of the NIRSPEC spectrum of 2M1510B obtained on 2017 August 2 (UT). Note that this model includes only one stellar component parameterized by effective temperature teff, surface gravity logg, radial velocity rv and rotational velocity vsini. The radial velocities shown do not take into account the barycentric motion at this epoch (-23-131 km/s).

Supplementary Table 2: Parameter values for fits to NIRSPEC observations of 2M1510A

| HJD – 2,400,000 | $T_{eff,1}$ (K) | $log_{10}g_1$ (cgs) | $RV_1$ (km/s) | $RV_2$ (km/s) | $f_2$ |
|---|---|---|---|---|---|
| 57967.73544 | 2754+17-33 | 5.20+0.13-0.17 | -1.9+0.5-0.7 | -25.6+0.5-0.6 | 0.77+0.05-0.05 |
| 58120.18151 | 2721+32-33 | 4.98+0.14-0.18 | -18.5+0.6-0.8 | -8.4+0.6-0.6 | |
| 58208.01057 | 2800+31-38 | 5.37+0.09-0.20 | -29.0+0.8-0.6 | 3.7+0.7-0.6 | 0.90+0.07-0.06 |
| 58261.89557 | 2752+28-20 | 5.18+0.16-0.08 | -5.1+0.6-0.5 | -19.7+0.7-0.7 | |
| 58266.90143 | 2745+27-28 | 5.09+0.10-0.16 | -18.9+0.6-0.7 | -5.6+0.7-0.5 | |
| 58271.85556 | 2759+28-27 | 5.32+0.13-0.20 | -29.5+0.6-0.6 | 5.0+0.7-0.7 | 0.94+0.05-0.05 |
| 58272.89590 | 2771+32-32 | 5.00+0.29-0.37 | -28.2+0.7-0.8 | 3.7+0.7-0.8 | 0.95+0.04-0.06 |
| Mean | 2755±20 | 5.17±0.12 | | | 0.88±0.07 |

Supplementary Table 3: Parameter values for fits to NIRSPEC observations of 2M1510B

| HJD – 2,400,000 | $T_{eff}$ (K) | $log_{10}g$ (cgs) | RV (km/s) | vsini (km/s) |
|---|---|---|---|---|
| 57967.25104 | 2705+35-24 | 5.07+0.21-0.17 | -12.4+0.8-0.7 | 14.1+1.2-1.3 |
| 58207.53513 | 2587+36-33 | 4.84+0.16-0.17 | -11.6+0.9-1.0 | 15.5+1.1-1.3 |
| 58261.42318 | 2654+34-33 | 4.85+0.19-0.19 | -11.7+0.9-1.0 | 15.2+1.4-1.2 |
| Mean | 2657±51 | 4.94±0.11 | -12.0±0.3 | 14.9±0.6 |

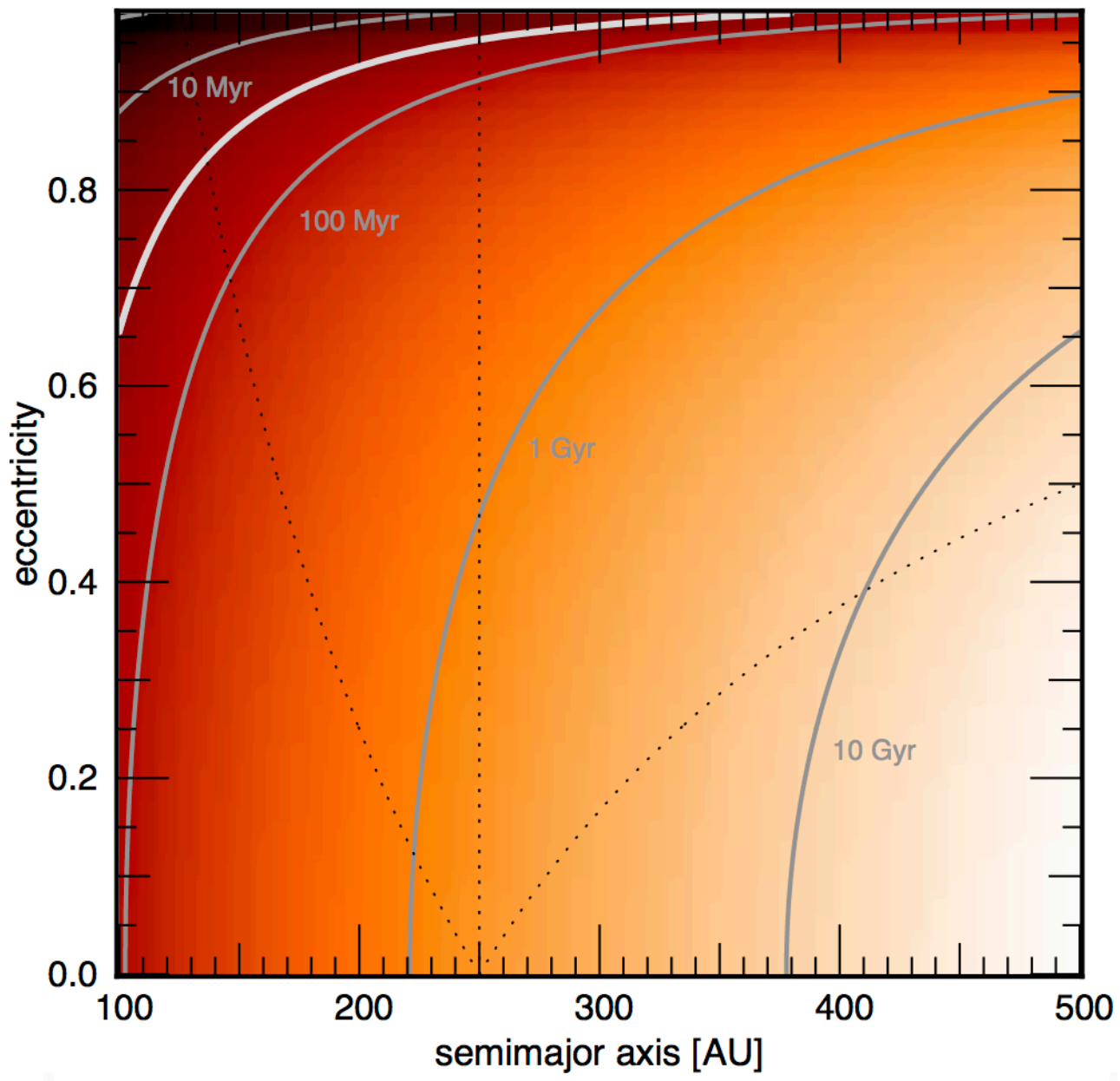

Supplementary Figure 9. The Lidov-Kozai timescale $\tau_{LK}$ (colour scale), as a function of orbital eccentricity and semi-major axis of 2M1510B about the 2M1510Aab binary. Grey lines give contours of constant $\tau_{LK}$, with the thickest, lightest contour corresponding to the system's age. The vertical dotted line corresponds to the projected separation of 2M1510A and 2M1510B, while the curved lines estimate the pair's semi-major axis should the current position be at apastron (left curve), or periastron (right curve). The former case has $\tau_{LK} < 45$ Myr for $e_{out} > 0.80$, in which case Lidov-Kozai oscillations could occur within the age of the system.

Supplementary Table 4: additional parameters resulting from our fit to the photometric and radial-velocimetric data.

| Parameters | Description | Posterior mean and 1σ confidence region | Unit |
|---|---|---|---|
| System parameters | | | |
| $(R_1 + R_2) / a$ | Scaled sum of radii | $0.0232_{-0.0011}^{+0.0011}$ | |
| $R_2/R_1$ | Radius ratio | 1.0 (fixed) | |
| $f_2/f_1$ | Surface Brightness ratio | 0.8 (fixed) | |
| sqrt(e) sin w | Eccentricity parameter | $-0.555_{-0.020}^{+0.020}$ | |
| sqrt(e) cos w | Eccentricity parameter | $0.0006_{-0.0321}^{+0.0322}$ | |
| $K_1$ | Primary semi-amplitude | $17.02_{-0.61}^{+0.64}$ | km/s |
| $K_2$ | Secondary semi-amplitude | $17.35_{-0.55}^{+0.54}$ | km/s |
| cos i | Cosine of sky inclination | $0.0266_{-0.0018}^{+0.0019}$ | |
| $\gamma_{\mathrm{NIRSPEC}}$ | Systemic velocity, NIRSPEC | $-12.87_{-0.40}^{+0.41}$ | km/s |
| $\gamma_{\mathrm{UVES}}$ | Systemic velocity, UVES | $-12.94_{-0.43}^{+0.43}$ | km/s |
| Nuisance parameters | | | |
| $\sigma_{1,\mathrm{NIRSPEC}}$ | Jitter, primary, NIRSPEC | $0.9_{-0.5}^{+0.7}$ | km/s |
| $\sigma_{2,\mathrm{NIRSPEC}}$ | Jitter, secondary, NIRSPEC | $0.9_{-0.5}^{+0.8}$ | km/s |
| $\sigma_{1,\mathrm{UVES}}$ | Jitter, primary, UVES | $2.0_{-0.6}^{+0.8}$ | km/s |
| $\sigma_{2,\mathrm{UVES}}$ | Jitter, secondary, UVES | $0.4_{-0.3}^{+0.5}$ | km/s |

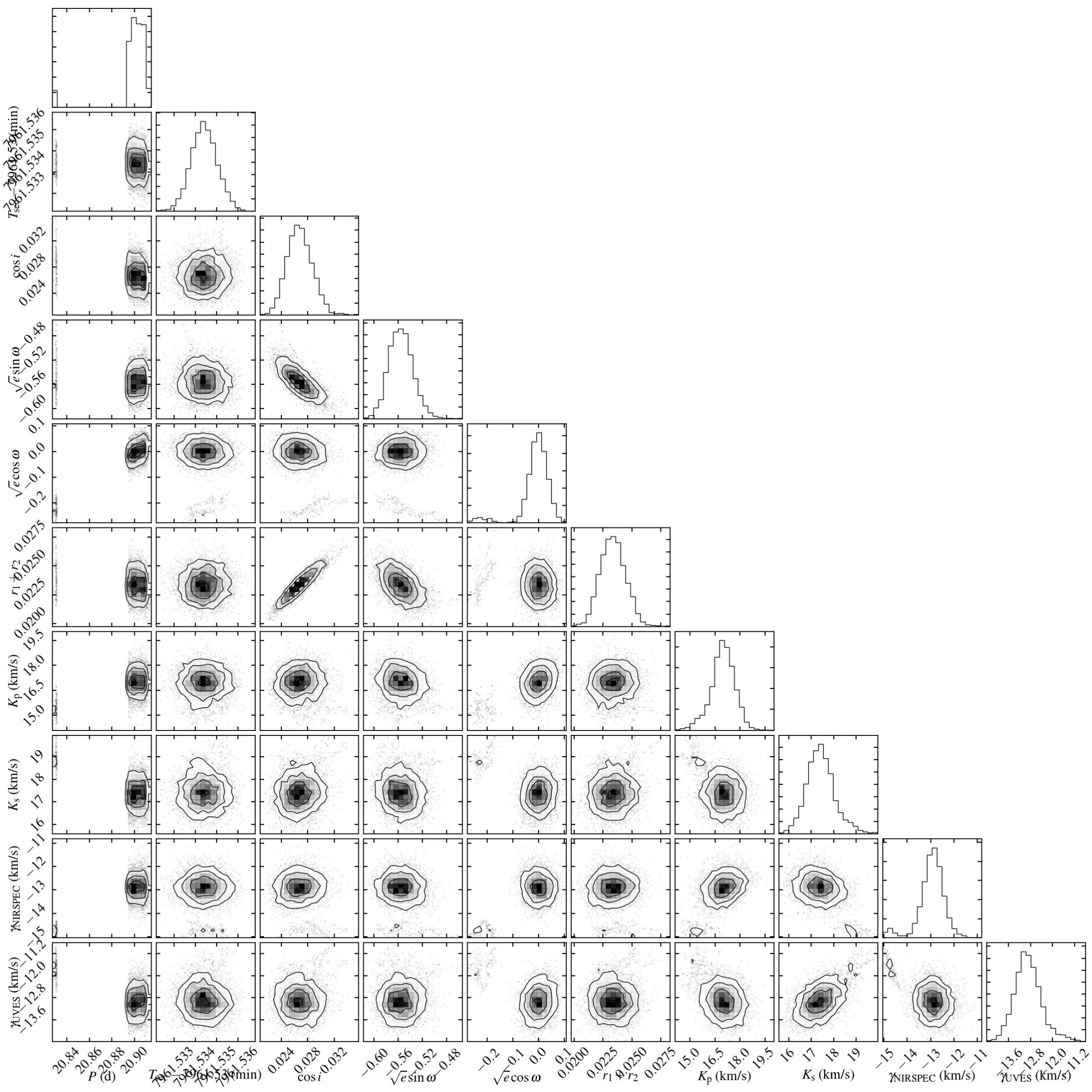

Supplementary Figure 10. Corner plot showing the posterior parameter distributions of our joint MCMC orbital model fit. The panels along the right diagonal display the marginalized distributions of each parameter, the inner panels show two-parameter correlations with all other parameters marginalized. The parameters shown are the period P (in days), the time of eclipse (in heliocentric Julidan data JD – 2450000), the cosine of inclination *i*, the factors $\sqrt{e}$ sin ω and $\sqrt{e}$ cos ω where e is eccentricity and ω the argument of periapsis, the sum of the fractional component radii ($r_1$+$r_2$ = ($R_1$+$R_2$)/a), the primary velocity amplitude, the secondary velocity amplitude, and velocity offsets (including systemic motion) for NIRSPEC ($\gamma_{NIRSPEC}$) and UVES ($\gamma_{UVES}$) observations.

Supplementary Table 5: list of double-line systems shown in Figure 2, along with their parameters and references.

| Object | Mass | Radius | Age | Reference |
|---|---|---|---|---|
| 2M0535-05 A | 0.0572 $M_\odot$ | 0.690 $R_\odot$ | 1-2 Myr | [11] |
| 2M0535-05 B | 0.0366 $M_\odot$ | 0.540 $R_\odot$ | 1-2 Myr | [11] |
| EPIC 2037103875A | 0.1183 $M_\odot$ | 0.417 $R_\odot$ | 5-10 Myr | [73] |
| EPIC 2037103875B | 0.1076 $M_\odot$ | 0.450 $R_\odot$ | 5-10 Myr | [73] |
| 2MJ0446+19B | 0.19 $M_\odot$ | 0.21 $R_\odot$ | 125 Myr | [73] |
| MH0 9B | 0.172 $M_\odot$ | 0.321 $R_\odot$ | 125 Myr | [73] |
| AD 2615 A | 0.212 $M_\odot$ | 0.233 $R_\odot$ | 600-800 Myr | [73] |
| AD 2615 B | 0.255 $M_\odot$ | 0.267 $R_\odot$ | 600-800 Myr | [73] |
| AD 3814B | 0.2022 $M_\odot$ | 0.2256 $R_\odot$ | 600-800 Myr | [73] |
| LP107-25B | 0.1518 $M_\odot$ | 0.1836 $R_\odot$ | | [74] |
| WTS19g-4 B | 0.143 $M_\odot$ | 0.174 $R_\odot$ | | [75] |
| LP661-13B | 0.1940 $M_\odot$ | 0.2174 $R_\odot$ | | [75] |
| HATS551-027B | 0.179 $M_\odot$ | 0.216 $R_\odot$ | | [76] |

Supplementary Table 6: List of single-line systems shown in Figure 2, along with their parameters and references.

| Object | Mass | Radius | Reference |
|---|---|---|---|
| LP261-75 b | 0.0650 $M_\odot$ | 0.0923 $R_\odot$ | [75] |
| CoRoT-3b | 21.23 $M_{Jup}$ | 0.993 $R_{Jup}$ | [77] |
| NLTT41135b | 33.7 $M_{Jup}$ | 1.13 $R_{Jup}$ | [77] |
| CoRoT-15b | 63.3 $M_{Jup}$ | 1.12 $R_{Jup}$ | [77] |
| LHS6343C | 62.1 $M_{Jup}$ | 0.783 $R_{Jup}$ | [77] |
| Kepler-39b | 19.1 $M_{Jup}$ | 1.11 $R_{Jup}$ | [77] |
| KELT 1b | 27.38 $M_{Jup}$ | 1.116 $R_{Jup}$ | [77] |
| KOI-205b | 40.8 $M_{Jup}$ | 0.82 $R_{Jup}$ | [77] |
| KOI-415b | 62.14 $M_{Jup}$ | 0.79 $R_{Jup}$ | [77] |
| KOI-189b | 78.0 $M_{Jup}$ | 0.998 $R_{Jup}$ | [77] |
| CoRoT-33b | 59.0 $M_{Jup}$ | 1.10 $R_{Jup}$ | [77] |
| EPIC 201702477b | 66.9 $M_{Jup}$ | 0.757 $R_{Jup}$ | [77] |
| WASP-30b | 62.5 $M_{Jup}$ | 0.951 $R_{Jup}$ | [13] |
| EBLM J1219-39b | 95.4 $M_{Jup}$ | 1.142 $R_{Jup}$ | [13] |
| EBLM J0113+31b | 0.186 $M_\odot$ | 0.209 $R_\odot$ | [78] |
| EBLM J0055-57b | 85.2 $M_{Jup}$ | 0.84 $R_{Jup}$ | [60] |
| EBLM J0543-56b | 0.164 $M_\odot$ | 0.193 $R_\odot$ | [79] |
| EBLM J0954-23b | 0.098 $M_\odot$ | 0.10 $R_\odot$ | [79] |
| EBLM J1013+01b | 0.177 $M_\odot$ | 0.215 $R_\odot$ | [79] |
| EBLM J1038-37b | 0.174 $M_\odot$ | 0.21 $R_\odot$ | [79] |
| EBLM J1115-36b | 0.179 $M_\odot$ | 0.193 $R_\odot$ | [79] |
| EBLM J1403-32b | 0.276 $M_\odot$ | 0.283 $R_\odot$ | [79] |
| EBLM J1431-11b | 0.121 $M_\odot$ | 0.149 $R_\odot$ | [79] |
| EBLM J2017+02b | 0.136 $M_\odot$ | 0.15 $R_\odot$ | [79] |
| EBLM J0218-31b | 0.21 $M_\odot$ | 0.27 $R_\odot$ | [80] |
| EBLM J2308-46b | 0.22 $M_\odot$ | 0.21 $R_\odot$ | [80] |
| EBLM J2349-32b | 0.11 $M_\odot$ | 0.16 $R_\odot$ | [80] |
| OGLE-TR-106 b | 0.116 $M_\odot$ | 0.181 $R_\odot$ | [81] |
| OGLE-TR-122 b | 0.092 $M_\odot$ | 0.12 $R_\odot$ | [81] |
| OGLE-TR-123 b | 0.085 $M_\odot$ | 0.133 $R_\odot$ | [81] |
| RR-Cae b | 0.1825 $M_\odot$ | 0.2090 $R_\odot$ | [81] |
| KIC 1571511 b | 0.141 $M_\odot$ | 0.1783 $R_\odot$ | [81] |
| HAT-TR-205-013 b | 0.124 $M_\odot$ | 0.167 $R_\odot$ | [81] |
| SDSS 01380016 b | 0.132 $M_\odot$ | 0.165 $R_\odot$ | [81] |
| SDSS 0857+0342 | 0.087 $M_\odot$ | 0.1096 $R_\odot$ | [81] |
| GK-Vir b | 0.116 $M_\odot$ | 0.155 $R_\odot$ | [81] |
| C4780 Bb | 0.096 $M_\odot$ | 0.104 $R_\odot$ | [81] |
| HATS551-021 b | 0.132 $M_\odot$ | 0.154 $R_\odot$ | [81] |
| HATS551-019 b | 0.17 $M_\odot$ | 0.18 $R_\odot$ | [81] |
| HATS551-016 b | 0.110 $M_\odot$ | 0.147 $R_\odot$ | [81] |
| Proxima | 0.123 $M_\odot$ | 0.141 $R_\odot$ | [82] |
| RIK 72 b | 56.1 $M_{Jup}$ | 3.06 $R_{Jup}$ | [7] |
| WASP-128 b | 37.5 $M_{Jup}$ | 0.94 $R_{Jup}$ | [14] |

Supplementary Table 7: List of young triple systems.

| System | Notes | references |
|---|---|---|
| DENIS 0205-1159ABC | Claimed from elongated PSF | [83,84] |
| 2MASS 0838+15ABC | Fully resolved | [85] |
| VHS 1256-1257ABc | Fully resolved | [86, 87] |
| 2MASS 0920+3517ABC | Not resolved, but highly likely | [88] |
| 2MASS 0700+3157ABC | Unverified, maybe not young | [88] |
| 2MASS 0249-0557ABc | Fully resolved | [89] |

References for the Supplementary Information: